\documentclass[prl,twocolumn,aps,epsfig,prbbib]{revtex4-1}
\usepackage{graphicx}

\begin{document}

\title{Tuning the properties of complex transparent conducting oxides:
role of crystal symmetry, chemical composition and carrier generation}

\author{Julia E. Medvedeva}\email{juliaem@mst.edu}
\author{Chaminda L. Hettiarachchi}

\affiliation{Department of Physics, Missouri University 
of Science \& Technology, Rolla, MO 65409, USA}

\begin{abstract}
The electronic properties of single- and multi-cation transparent conducting oxides (TCOs) are investigated using first-principles density functional approach.
A detailed comparison of the electronic band structure of 
stoichiometric and oxygen deficient In$_2$O$_3$, $\alpha$- 
and $\beta$-Ga$_2$O$_3$, rock salt and wurtzite ZnO, and layered InGaZnO$_4$
reveals the role of the following factors which
govern the transport and optical properties of these TCO materials: 
(i) the crystal symmetry of the oxides, including both the oxygen coordination 
and the long-range structural anisotropy;
(ii) the electronic configuration of the cation(s), specifically,
the type of orbital(s) -- $s$, $p$ or $d$ -- which form the conduction band; 
and (iii) the strength of the hybridization
between the cation's states and the $p$-states of the neighboring oxygen atoms.
The results not only explain the experimentally observed trends in the electrical 
conductivity in the single-cation TCO, but also demonstrate that multicomponent oxides 
may offer a way to overcome the electron localization bottleneck which limits
the charge transport in wide-bandgap main-group metal oxides.
Further, the advantages of aliovalent substitutional doping -- 
an alternative route to generate carriers in a TCO host -- 
are outlined based on the electronic band structure   
calculations of Sn, Ga, Ti and Zr-doped InGaZnO$_4$. We show that 
the transition metal dopants offer a possibility to improve
conductivity without compromising the optical transmittance.


\end{abstract}


\maketitle


\subsection{I. Introduction}

Multicomponent transparent conducting oxides (TCOs) -- complex oxides 
which contain several types of main group metal ions -- 
have been developed \cite{Shannon,Chopra,Dawar,MRS,spinel-review,Mason-review} 
to broaden the range of transparent conducting and semiconducting materials
required for a variety of applications, including
photovoltaic cells, flat panel displays, and flexible and invisible electronics.
Binary and ternary compounds and solid solutions with electrical,
optical
and mechanical
properties controlled via chemical composition,
have been a subject of numerous investigations, e.g.,
see \cite{Chopra,Dawar,MRS,spinel-review,Mason-review} for reviews.
Since the 1990s, the multi-cation TCOs which include metal ions
beyond the traditionally employed Sn, Cd, In and Zn have emerged, for example,
MgIn$_2$O$_4$ \cite{Unno}, GaInO$_3$ \cite{Phillips} and
the so-called 2-3-3 or 3-3-4 systems where
the numbers correspond to divalent, trivalent and tetravalent
cations \cite{Freeman}. Among them, layered compounds of the homologous series (In,Ga)$_2$O$_3$(ZnO)$_n$, 
where $n$=integer, attract an increasing attention 
\cite{Orita-exp,Science,Nature,anisotropy,Kazeoka,thermoelectr,NomuraIGZO,dasilva,bandstrIGZO,Cho,Lim,WeiPRB,amorphTCO}, 
originally, due to a possibility to enhance conductivity 
via spatial separation of the carrier donors (dopants) located in insulating layers and the conducting layers which transfer the carriers effectively, i.e., without charge scattering.
However, 
the electrical conductivity and carrier mobility observed to-date 
in these complex materials, $\sigma$=100-400 S/cm and $\mu$=10-20 cm$^2$/V$\cdot$s, respectively
\cite{Orita-exp,Nature,thermoelectr,Hiramatsu,Moriga,PhilMag} 
are considerably lower than those achieved in the single-cation TCOs, such as In$_2$O$_3$, SnO$_2$ or ZnO, with $\sigma$=10$^3$-10$^4$ S/cm and $\mu$=50-100 cm$^2$/V$\cdot$s.

The low conductivity in (In,Ga)$_2$O$_3$(ZnO)$_n$ which decreases continuously as the number of ZnO layers, $n$, increases, 
along with the observed strong anisotropy of the transport properties in these complex oxides \cite{anisotropy,thermoelectr}  
was assumed to correlate with the density of the octahedrally coordinated In atoms that changes with $n$. 
Since late 1970's, the octahedral oxygen coordination of cations was believed
to be essential for a good transparent conductor 
\cite{Shannon,Mason-review,Freeman,Li-implant,Woodward}.
Accordingly, it has been suggested that in (In,Ga)$_2$O$_3$(ZnO)$_n$ the charge is transferred only within 
the octahedrally coordinated In layers while the Ga-Zn layers where both cations have tetrahedral oxygen coordination, were believed to be non-conducting \cite{Orita-exp,cluster}.
However, the above assumption appears to be in conflict with:

(i) experimental observations that the mobility and conductivity 
are independent of the compositional variations in the {\it amorphous} InGaO$_3$(ZnO)$_n$ 
and the conclusion that this material is a Zn 4$s$ conductor
\cite{PhilMag},

(ii) electronic band structure investigations which showed that 
all chemically and structurally distinct layers in InGaZnO$_4$ ($n$=1) 
and in similar compounds are expected 
to participate in the charge transport owing to the hybrid nature of 
the bottom of the conduction band formed from the $s$-states of {\it all} cations 
in the cell and the $p$-states of their oxygen neighbors \cite{my2epl}.

The above discrepancy calls for further microscopic analysis of the layered
oxides. 
In particular, the underlying mechanisms associated with carrier generation 
in the complex TCO materials should be addressed. 
The purpose of this work is not only to determine the properties of oxygen deficient 
or substitutionally doped InGaZnO$_4$
but also to understand how the presence of several cations 
of different valence, ionic size, electronic configuration and oxygen coordination 
affects the carrier generation and the overall electronic and optical 
properties of complex TCOs.

For the materials of the (In,Ga)$_2$O$_3$(ZnO)$_n$ family, it has been accepted 
that free carriers appear due to the formation of oxygen vacancies. 
Also, it was assumed that each vacancy donates two extra electrons, 
i.e., there is no charge compensation via natural defects 
or defect complexes. 
However, it was shown theoretically \cite{Hosono-defects-vasp} 
that uncompensated oxygen vacancy in InGaZnO$_4$ results in a deep level within 
the band gap and, thus, free carriers cannot be efficiently generated. 
Therefore, oxygen vacancy cannot explain the observed conducting behavior 
in this material.

Substitutional doping, an alternative route to generate carriers, 
remains unexplored in InGaZnO$_4$ and the compounds of the homologous series. 
One of the main reasons for this is the structural and compositional complexity of multicomponent oxides.
Indeed, the presence of several cations with different valence (e.g., In$^{3+}$ and Ga$^{3+}$ 
vs Zn$^{2+}$) and oxygen coordination (octahedral for In vs tetrahedral for Ga and Zn)
may facilitate a non-uniform distribution of impurities
including their clustering or secondary phase formation.
Furthermore, targeted doping required for extra carrier generation
via aliovalent substitution, becomes difficult as the number of 
the structurally and/or chemically distinct cations in the complex material increases.
The aliovalent substitutional dopants known to be efficient carrier donors 
in the constituent single-cation TCO, e.g., Al$^{3+}$ in ZnO,
may not be suitable for the complex oxides since they may prefer to substitute
the cations of the same valence (In$^{3+}$ or Ga$^{3+}$). 
In this case, extra carriers are not generated. Formation of charge compensating defects and defect complexes is also more likely in complex oxides. 
Thus, the range of dopants which are efficient for a given multicomponent material may be quite narrow.

In this work, first-principles density functional approach is employed to 
investigate possible carrier generation mechanisms in the layered InGaZnO$_4$ ($n$=1) 
and its constituent single-cation oxides, In$_2$O$_3$, Ga$_2$O$_3$ and ZnO, 
and to understand how each of the mechanisms
affects the resulting electronic and optical properties.
First, we perform electronic band structure calculations of oxygen deficient materials
to understand the origin of the observed anisotropic transport properties
in the layered multicomponent oxide. For this, we analyze the role of crystal symmetry 
and metal-oxygen hybridization by comparing the properties of 
high- and low-symmetry oxide phases. In addition, our comparative studies of several 
oxides allow us to understand how chemical composition (In vs Zn vs Ga oxides)
affects the resulting optical and electronic properties.

Next, in search of efficient substitutional dopants, we study the structural and 
electronic properties of InGaZnO$_4$ doped with either conventional dopants 
such as the post-transition metal ions (Sn, Ga) or with the transition metals (Ti, Zr). 
We find that the optical and transport properties are both sensitive to the ionic radius 
of the dopant and its electronic configuration.
In particular, we show that the presence of the localized $d$-states is beneficial 
for balancing good conductivity with low optical absorption.
Based on the results obtained, the (dis)advantages of both carrier generation 
mechanisms -- oxygen reduction and aliovalent substitutional doping -- are discussed.
We conclude that the conductivity currently achieved in the oxygen deficient 
(In,Ga)$_2$O$_3$(ZnO)$_n$ can be significantly improved with proper doping of the host 
material.

The paper is organized as follows. First, theoretical approaches 
and structural peculiarities of the compounds under investigation 
are outlined in Section II.
Studies of oxygen deficient In$_2$O$_3$, $\alpha$- and $\beta$-Ga$_2$O$_3$, 
cubic and hexagonal ZnO  are presented in Section III.
Here we discuss the formation energies of the oxygen vacancy 
in various charge states and 
the changes in the host electronic band structure caused by 
the presence of the oxygen defect. 
Significantly, the range of materials investigated allows us to address 
the long-standing question about the role of the crystal symmetry, 
both local (oxygen coordination) and long-range (structural anisotropy), 
in the charge transport.
In addition, we compare the electronic band structure features of the three 
oxides and 
draw general conclusions about the role of the cation-anion orbital hybridization 
in the resulting transport properties. In Section IV, we perform similar 
investigations for the layered undoped stoichiometric and oxygen deficient 
InGaZnO$_4$.

In Section V, aliovalent substitutional doping is considered. 
For each of the dopants studied, i.e., Sn, Ga, Ti or Zr, several distinct site locations 
in the layered InGaZnO$_4$ are compared based on 
the total energy calculations. The results, along with the considerations
of the dopant and the substituted cation valences, allow us to judge 
the efficiency of each dopant, i.e., its capability 
to donate extra electrons. 
Further in this section, the role of the dopant 
electronic configuration in both the transport and optical 
properties is discussed and practical ways to optimize the former without 
compromising the latter are outlined. 
Finally, calculations for InGaZnO$_4$ with interstitial oxygen atoms located 
within either the In-O or Ga-Zn-O layers are performed to investigate 
a possibility of charge compensation in substitutionally doped materials. 

We discuss the implications of the obtained results in Section VI.

\subsection{II. Approach}

First-principles full-potential linearized augmented plane wave method
\cite{FLAPW,FLAPW1}
with the local density approximation (LDA) is employed for the electronic band structure
investigations. 
Cutoffs for the basis functions, 16.0 Ry, and potential representation, 81.0 Ry,
and expansion in terms of spherical harmonics with $\ell \le$ 8 inside the muffin-tin spheres
were used.
The muffin-tin radii are 2.3 for In; 1.85 a.u. for Ga; 1.9 for Zn; and 1.6 a.u. for O atoms.
Summations over the Brillouin zone were carried out
using at least 23 special {\bf k} points in the irreducible wedge.
For every structure investigated, the internal positions of atoms 
were optimized via the total energy and atomic forces minimization.

To study the electronic properties of undoped stoichiometric oxides,
we also employed the self-consistent screened-exchange LDA (sX-LDA) 
method \cite{sxlda,sxlda1,sx-Asahi,sx-paper,sx-Kim} for more accurate 
description of the band gap values and the valence/conduction band edges. 
The sX-LDA is known to provide a better energy functional beyond LDA or 
generalized gradient approximation (GGA) by modeling the exchange-correlation 
hole within the {\it nonlocal} density scheme.

InGaZnO$_4$ has rhombohedral $R\bar{3}m$ crystal structure of YbFe$_2$O$_4$ type 
\cite{str,str-all}. Indium atoms reside in the 3(a) positions (Yb), while
both Ga$^{3+}$ and Zn$^{2+}$ atoms are in the 6(c) positions (Fe) and
are distributed randomly \cite{Li} -- as confirmed by our total energy 
calculations \cite{my2epl}. Because of the different ionic radii, the Ga and Zn atoms
have different $z$ component of the internal site position (0,0,$z$): we find
$z$=0.214 for Ga and $z$=0.218 for Zn.
The optimized cation-anion distances are 2.16--2.22 \AA \, for In-O;
1.88 \AA \, and 1.96--2.16 \AA \, for the Ga-O in the $ab$-plane and 
along the [0001] direction, respectively; and 
1.97 \AA \, and 1.99--2.36 \AA \, for the planar and apical Zn-O distances, respectively.
These values correlate with the ionic radii of the cations (0.94 \AA \, for six-coordinated
In, 0.61 \AA \, for four-coordinated Ga and 0.74 \AA \, for four-coordinated Zn)
and correspond to the cation-anion distances in the single-cation oxides, i.e., 
In$_2$O$_3$ (2.13-2.25 \AA), Ga$_2$O$_3$ (1.83-2.07 \AA) 
and ZnO (1.97-2.14 \AA).

To model isolated point defects, the following supercells were constructed:
a 49-atom supercell with 
the lattice vectors (30$\bar{2}$), ($\bar{1}$12) and (02$\bar{1}$), given 
in the units of the rhombohedral primitive cell vectors, and a 77-atom supercell with 
the lattice vectors (30$\bar{2}$), ($\bar{1}$3$\bar{1}$) and ($\bar{2}$12)
for InGaZnO$_4$;
a 40-atom and 80-atom supercells for bixbyite In$_2$O$_3$;
a 120-atom supercell for monoclinic $\beta$-Ga$_2$O$_3$;
a 80-atom supercell for rhombohedral (corundum) $\alpha$-Ga$_2$O$_3$;
a 84-atom supercell with the lattice vectors ($\bar{2}\bar{2}$1), 
($\bar{1}$31) 
and (2$\bar{3}$1), given in the units of the hexagonal primitive cell vectors 
for wurtzite ZnO; and 64 and 128-atom supercells for cubic ZnO.

To compare the properties of oxides, we choose the supercells which result in 
a similar defect concentration, namely, 0.8-1.0$\times$10$^{21}$cm$^{-3}$. 
This corresponds to the distance between the oxygen defects of 8-10 \AA.
We stress that such high concentrations may not be experimentally feasible  
in all materials under consideration. However, our goal is (i) to compare the properties 
in oxides under the same (perhaps, hypothetical) conditions, and (ii) to demonstrate 
that even when the defects are relatively close to each other in our model and 
so their wavefunctions are expected to overlap significantly and, hence, form a broad band,
we obtain a localized state (as, for example, in $\beta$-Ga$_2$O$_3$, see below).

The oxygen vacancy with three charge states, i.e., neutral V$_O^0$ 
and ionized V$_O^{+}$ or V$_O^{2+}$,
was modeled using a corresponding
background charge. The defect formation energy is calculated as:
\begin{equation}
\Delta H(E_F,\mu) = E_{defect} - E_{host} + \sum \pm (\mu^0 + \Delta \mu ) + qE_F,
\end{equation}
where $E_{defect}$ and $E_{host}$ are the total energies for 
the oxygen deficient oxide and the stoichiometric oxide in the same-size 
supercell, respectively; $\Delta \mu$ is the chemical potential 
for atom added to (--) or removed from (+) the lattice; 
$q$ is the defect charge state and $E_F$ is the Fermi energy.
The chemical potentials $\Delta \mu$ are taken with respect to the LDA energy 
$\mu^0$ of the elementary metals or the O$_2$ molecule. We  
apply the thermodynamic stability condition for the oxides. 
Our calculated formation enthalpies, $\Delta H_f$,  
are in a good agreement with the experimental values 
given in parenthesis:
$\Delta H_f (In_2O_3)$=$-$8.2 eV ($-$9.6 eV); 
$\Delta H_f (ZnO)$=$-$3.4 eV ($-$3.6 eV); 
$\Delta H_f (Ga_2O_3)$=$-$9.7 eV ($-$11.2 eV); and
$\Delta H_f (InGaZnO_4)$=$-$11.97 eV (experimental data not available). 
The formation energies of substitutionally doped In$_2$O$_3$ and
InGaZnO$_4$ were calculated with respect to the bulk 
orthorhombic Ga, tetragonal In and Sn, and hexagonal Ti, Zr and Zn
metals.

\subsection{III. Electronic properties of In$_2$O$_3$, Ga$_2$O$_3$ and ZnO}

\subsubsection{III.1 Undoped stoichiometric TCO hosts}

\begin{figure*}
\includegraphics[width=17.0cm]{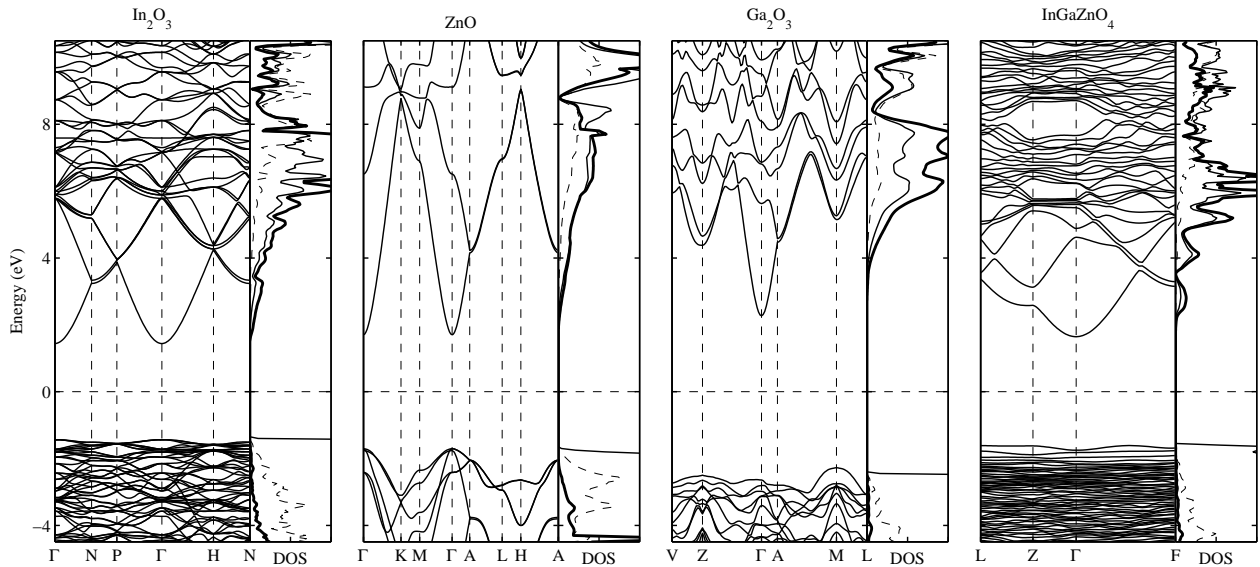}
\caption{Electronic band structure and density of states (DOS) of undoped 
stoichiometric (a) bixbyite In$_2$O$_3$, (b) wurtzite ZnO, 
(c) monoclinic Ga$_2$O$_3$, and (d) layered InGaZnO$_4$
as calculated within the self-consistent sX-LDA.
The thick, dashed and thin lines in the DOS plots represent the metal $s$, 
metal $p$ and oxygen $p$ states, respectively.}
\label{hosts}
\end{figure*}

Oxides of post-transition metals with $(n-1)d^{10}ns^2$ electronic 
configuration, such as In$_2$O$_3$, Ga$_2$O$_3$, ZnO and CdO 
share similar chemical, structural and electronic properties.
They have densely-packed structures with four- or six-coordinate metal ions,
Table \ref{table-hosts}.
Strong interactions between the oxygen $2p$ and metal $ns$ orbitals
give rise to electronic band structures which are qualitatively
similar among all of these oxides, Figure \ref{hosts}: the bonding and nonbonding O $2p$ states form the
valence band while the conduction band arises from the antibonding
M$s$-O$p$ interactions where M stands for metal ion. 
The empty $p$-states of the cation (dashed line) form the following band at a higher energy.
The partial density of states plots, Figure \ref{hosts}, reveal that the oxygen $2p$ (thin line)
and metal $ns$ (thick line) states make similar contributions to the conduction band.
This provides a three-dimensional M$s$-O$p$ network for charge transport
once extra carriers fill the band \cite{mybook}.

\begin{table*}
\begin{tabular}{lccccccc} \hline

& In$_2$O$_3$ & \multicolumn{2}{c}{ZnO} & \multicolumn{2}{c}{Ga$_2$O$_3$} & InGaZnO$_4$ \\ \hline

Lattice &   Bixbyite  & Wurtzite & Rocksalt & Monoclinic & Corundum & Rhombohedral \\

Coordination of cation & 6 & 4 & 6 & 6,4 & 6 & 6(In),4(Zn,Ga) \\

Coordination of anion & 4 & 4 & 6 & 4,3 & 4 & 4 \\

Fundamental gap, eV           
   & 2.88 (1.16) & 3.41 (0.81) & 2.49 (1.03) & 4.86 (2.32) & 5.16 (2.81) & 3.29 (1.19) \\

Optical (direct) band gap, eV 
   & 3.38 (1.52) & 3.41 (0.81) & 3.86 (2.28) & 4.91 (2.32) & 5.40 (3.06) & 3.40 (1.28) \\

Electron effective mass, in m$_e$  
   & 0.28 (0.18) & 0.35 (0.19) & 0.28 (0.22) & 0.34 (0.24) & 0.32 (0.23) & 0.34 (0.21) \\

Effective mass anisotropy, $\delta$ 
   & 1.00 (1.00) & 1.01 (1.25) & 1.00 (1.00) & 1.10 (1.23) & 1.00 (1.00) & 1.00 (0.97) \\ \hline 
\end{tabular}
\caption{Structural and electronic properties of TCO hosts -- the undoped stoichiometric oxides. 
The values are determined within the self-consistent screened-exchange local-density approximation and within LDA (given in parentheses). 
Anisotropy of the electron effective mass is defined as
$\delta=(m^{[100]}+m^{[010]})/2m^{[001]}$.}
\label{table-hosts}
\end{table*}

M$s$-O$p$ interactions result in a gap between the valence
and the conduction bands. In ZnO, the gap is direct
whereas in In$_2$O$_3$ and Ga$_2$O$_3$ the valence band maximum
is at the H point ([1$\bar{1}$1]) and M point ([$\bar{1}$11]), respectively, giving rise to indirect
band gap of 2.88 eV and 4.86 eV, respectively.
(For results on CdO, see Refs. \cite{my1epl,my-cdo1,my-cdo2,my-cdo3}.)
For TCO applications, direct optical band gaps are of primary importance.
The calculated values are given in Table \ref{table-hosts}. As one can see, 
our sX-LDA results are in excellent agreement with the reported experimental and
theoretical values,
namely, 2.9 eV for the fundamental band gap and 3.5-3.7 eV for the optical band gap 
for In$_2$O$_3$ \cite{Weiher,Hamberg,newgapIn2O3,bandstrIn2O3}, 
3.1-3.6 eV for ZnO \cite{gapZnO} and
4.5-4.9 eV for Ga$_2$O$_3$ \cite{gapGa2O3,gapGa2O3-2},
while LDA significantly underestimates the band gap values, as expected.

The M$s$-O$p$ overlap also determines the energy dispersion of 
the conduction band in these materials. 
Within the framework of {\bf k$\cdot$p} theory \cite{kp}, the electron effective mass 
can be found within the second-order perturbation:
\begin{equation}
\frac{m_e}{m_{ii}^{(c)}}=1+\frac{2}{m_e}\sum_{l\neq c} \frac{|\langle u^{(c)}|\hat{p}_i|u^{(l)}\rangle|^2}{E^{(c)}-E^{(l)}},
\label{kp-eq}
\end{equation}
where $\hat{\bf p}$ is the momentum operator, 
$|u^{(l)}\rangle$ is the Bloch wave function of the $l$'s band at the $\Gamma$ point
(wave vector {\bf k}=0) and $E^{(l)}$ is its energy.
Band label $c$ represents the conduction band, while the sum 
runs over all other bands.
In the oxides under consideration here, the electron effective mass is less than 
the mass of the electron, $m_e$. As it follows from Eq. \ref{kp-eq},
it is determined primarily by the valence band contributions ($E^{(l)}<E^{(c)}$), 
i.e., by the oxygen $2p$-states.
Thus, the network of alternating metal and oxygen atoms ensures
the small electron effective mass in the TCO hosts \cite{mybook}.

Moreover, the electron effective mass varies insignificantly with 
oxygen coordination which may be different for different phases of 
the same-cation oxide \cite{my2epl}. For example, for ZnO
in rock salt (octahedral coordination) or wurtzite (tetrahedral coordination)
structures, and for In$_2$O$_3$ in $Ia\bar{3}$ (bixbyite), $R\bar{3}c$ 
(corundum) or $I2_13$ structures, the effective masses vary by less than or about 15\%
(see also \cite{bandstrIn2O3}).
In addition, the effective mass remains nearly isotropic in all oxide phases considered
including those with well-defined long-range structural anisotropy. 
This is in accord with the investigations of complex oxides of main group 
metals \cite{my2epl,mycaalo} where isotropic electron effective mass is found 
for oxides with irregular atomic arrangements or large structural voids.
These observations explain the success of (i) amorphous TCOs whose 
electrical properties remain similar to those in the crystalline state, 
and (ii) TCO-based flexible transparent conducting coatings. 

As shown in the next section where we consider the conversion of the undoped stoichiometric TCO hosts 
from insulators into conductors, the M$s$-O$p$ origin of the conduction band plays critical role in determining the resulting transport properties.

\subsubsection{III.2 Oxygen deficient In$_2$O$_3$, ZnO and Ga$_2$O$_3$}

Oxides of post-transition metals exhibit a relatively low free energy 
of formation \cite{Reed} which favors large oxygen deficiencies 
even under equilibrium growth conditions.
This gives rise to the free-carrier densities as large as 
10$^{17}$-10$^{19}$ cm$^{-3}$ in In$_2$O$_3$ and 
ZnO \cite{Wit,Kroger,zunger-prl}.

\begin{figure*}
\includegraphics[width=8.0cm]{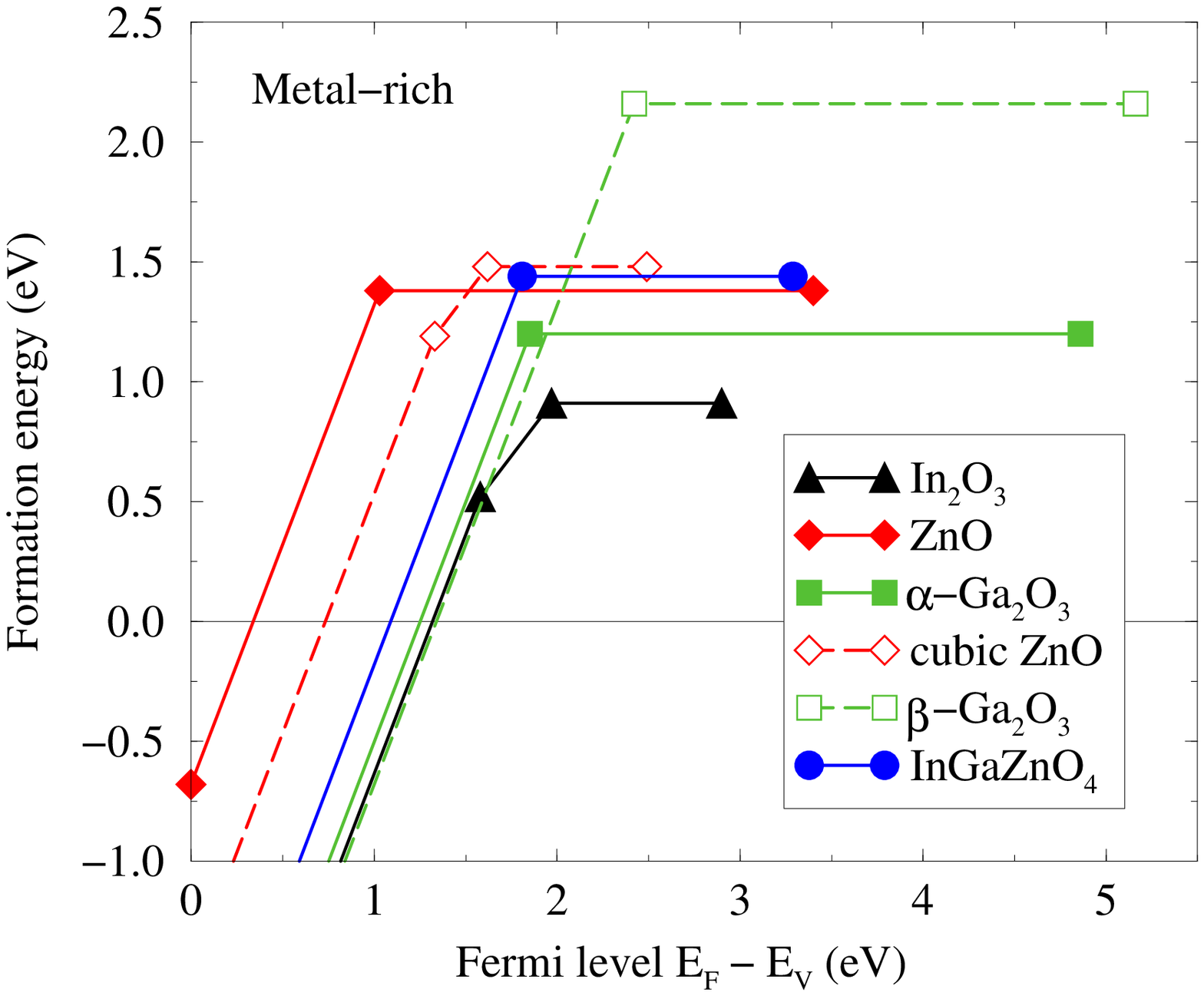}
\includegraphics[width=8.0cm]{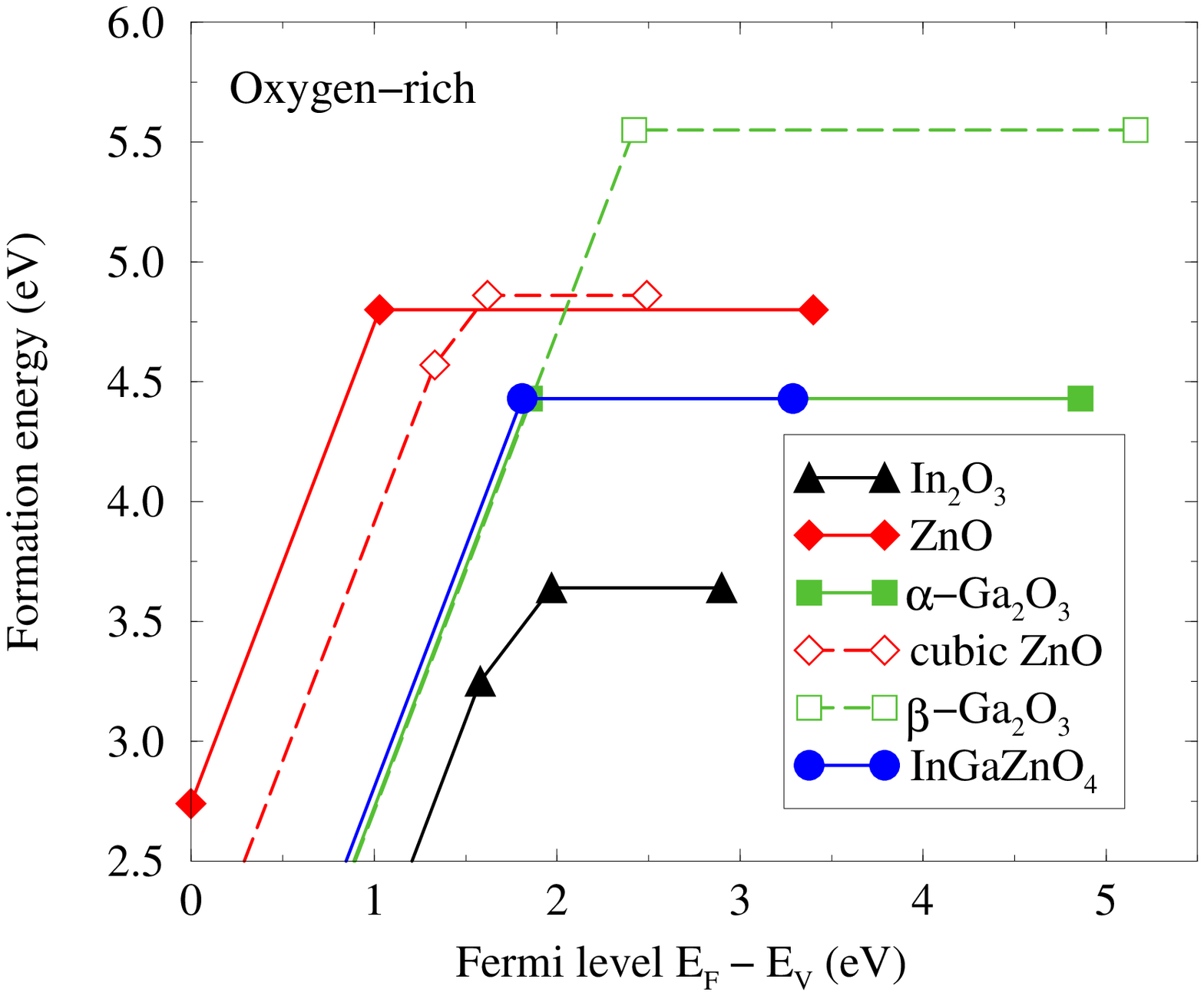}
\caption{Calculated formation energies of the oxygen vacancy 
as a function of the Fermi level. The symbols indicate 
the transition energies between different charge states of the oxygen defect.
Solid lines stand for In$_2$O$_3$ (triangles), wurtzite ZnO (diamonds), 
monoclinic $\beta$-Ga$_2$O$_3$ (squares) and InGaZnO$_4$ (circles).
Dashed lines are for rocksalt ZnO (open diamonds) and 
rhombohedral $\alpha$-Ga$_2$O$_3$ (open squares).
The oxygen defect concentration in all systems is 0.8-1.1$\times$10$^{21}$cm$^{-3}$. 
The results are obtained within LDA.}
\label{formation}
\end{figure*}

Removal of an oxygen atom from the metal oxide leaves two extra electrons 
in the crystal. 
Whether one or both of these electrons become mobile (free) carriers or
remain localized at the vacancy site is determined by the formation energy 
of the oxygen defect which can have various charge states
(i.e., V$_O^0$, V$_O^{+}$ or V$_O^{2+}$).
We find that in all oxides investigated in this work, the neutral
oxygen defect V$_O^0$ corresponds to the ground state 
of the oxygen vacancy, Fig. \ref{formation}.  
This is in accord with the results for In$_2$O$_3$ and ZnO 
by Lany and Zunger \cite{zunger-prl}.
From the electronic band structure calculations for oxygen deficient 
In$_2$O$_3$, Ga$_2$O$_3$ and ZnO, we find that the uncompensated 
oxygen vacancy, V$_O^0$, gives rise to a fully occupied 
(by the two vacancy-induced electrons), thus, non-conductive state, 
as the Fermi level falls into a gap between the defect state and 
the rest of the conduction band. 

The observed conductivity in In$_2$O$_3$ and ZnO was attributed to 
the photoexcited oxygen vacancy, i.e., V$_O^{1+}$, 
which provides a metastable conductive state \cite{zunger-prl}.
Indeed, when the oxygen vacancy is singly ionized (V$_O^{+}$), 
the lowest {\it single} conduction band associated with 
the presence of the oxygen defect becomes half occupied, 
leading to a conducting behavior \cite{fullycomp}. 
Within the local density approximation (LDA), we estimate the optical 
excitation V$_O^0 \rightarrow V_O^{+} + e^-$ 
to be 1.64 eV, 1.48 eV and 2.02 eV for In$_2$O$_3$, ZnO and Ga$_2$O$_3$, 
respectively. 
These values are underestimated because LDA underestimates 
the band gap by 2-3 eV, cf., Table \ref{table-hosts}.

Thus, only if the oxygen vacancy is excited to V$_O^{+}$, 
a conducting behavior may occur. 
The conductive state associated with V$_O^{+}$ is found to be located
closer to the conduction band minimum (CBM) as compared to 
V$_O^0$, Table \ref{table-vacancy}. 
Again, the values in Table \ref{table-vacancy} for the location of 
the defect states with respect to band edges may be underestimated,
however, we expect that LDA captures the trend
among the oxides \cite{mysXLDA} and below we use the LDA results 
to compare with those obtained 
for oxygen deficient $\alpha$-Ga$_2$O$_3$, rock salt ZnO and InGaZnO$_4$.

Despite similar defect concentration and, hence, similar 
distance between the oxygen defects (8-10 \AA) in 
In$_2$O$_3$, ZnO and $\beta$-Ga$_2$O$_3$,
the obtained widths of the lowest conduction band in the conductive oxides 
(i.e., those with V$_O^{+}$ defect) 
differ significantly, Table \ref{table-vacancy}.
The lower energy dispersion of the conduction band 
in $\beta$-Ga$_2$O$_3$ suggests that the extra electrons
tend to localize -- in marked contrast to the extended state in In$_2$O$_3$.
Accordingly, the calculated Fermi electron group velocities in the conductive oxides
differ by almost an order of magnitude, 
Table \ref{table-vacancy}. We stress that these differences cannot be accounted 
by the variation in the electron effective masses of stoichiometric 
materials (cf., Table \ref{table-hosts}).
The obtained trend in the electron group velocities is in accord 
with experimental results: indium oxide is known
to be the best TCO in terms of the minimum resistivity achieved, 
whereas $\beta$-Ga$_2$O$_3$ has not been considered as a viable TCO and, 
moreover, is believed to be the constituent detrimental for 
the charge transport of a multicomponent oxide. 
To understand the microscopic origin of the different electronic properties
of the oxygen deficient oxides, further comparative analysis will follow 
(see Sections III.3-5).

\begin{figure*}
\includegraphics[width=5.3cm]{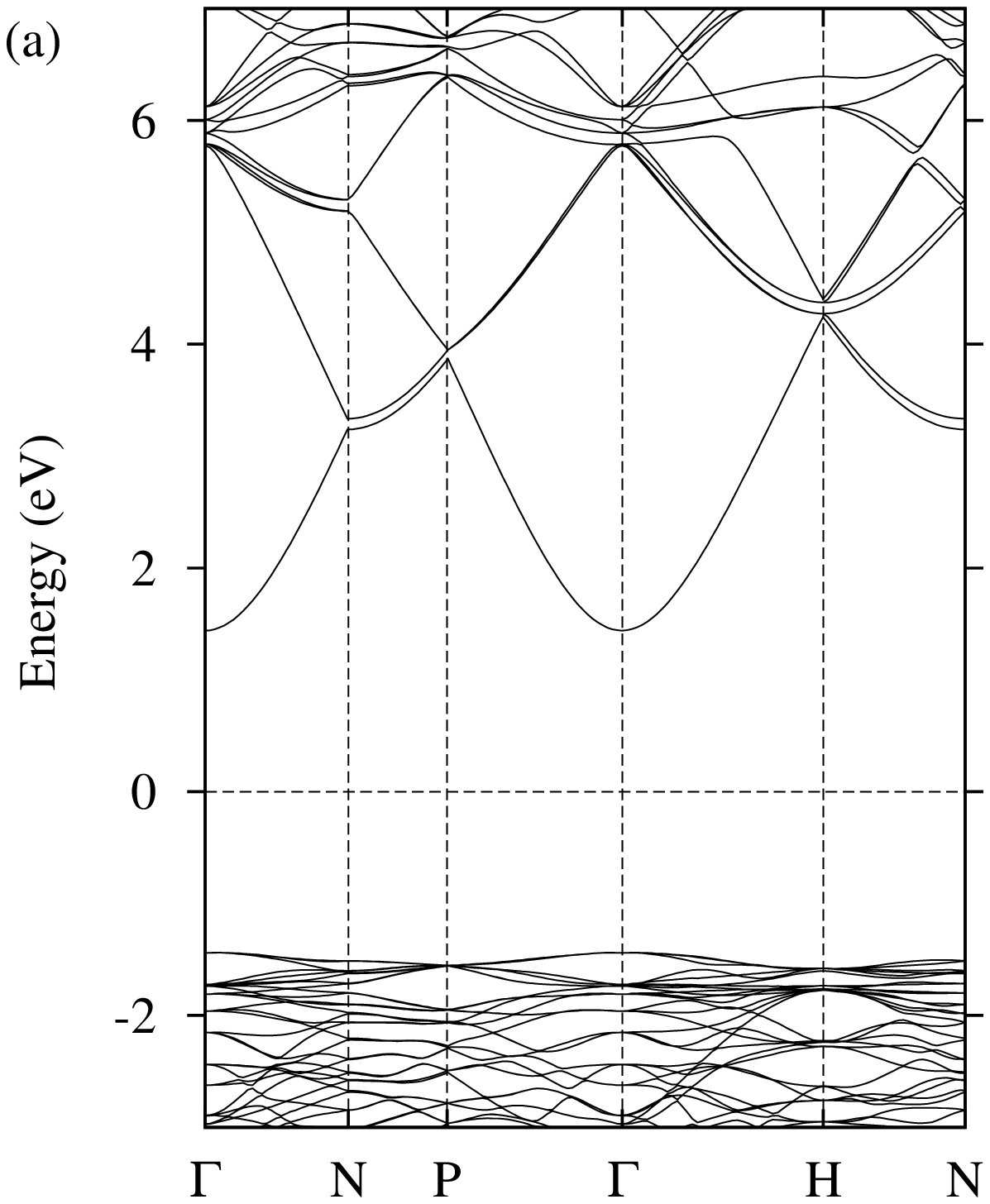}
\includegraphics[width=5.3cm]{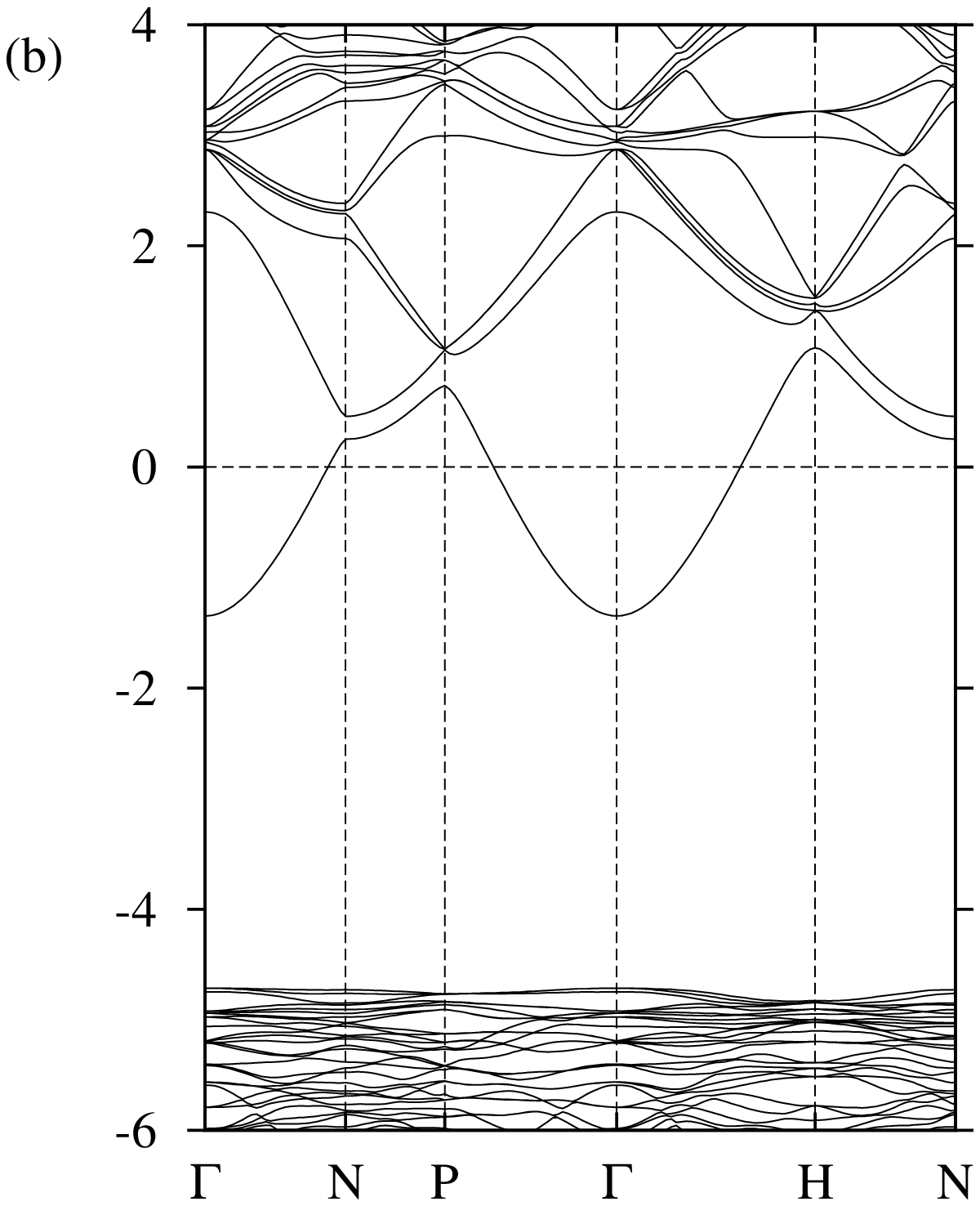}
\includegraphics[width=5.3cm]{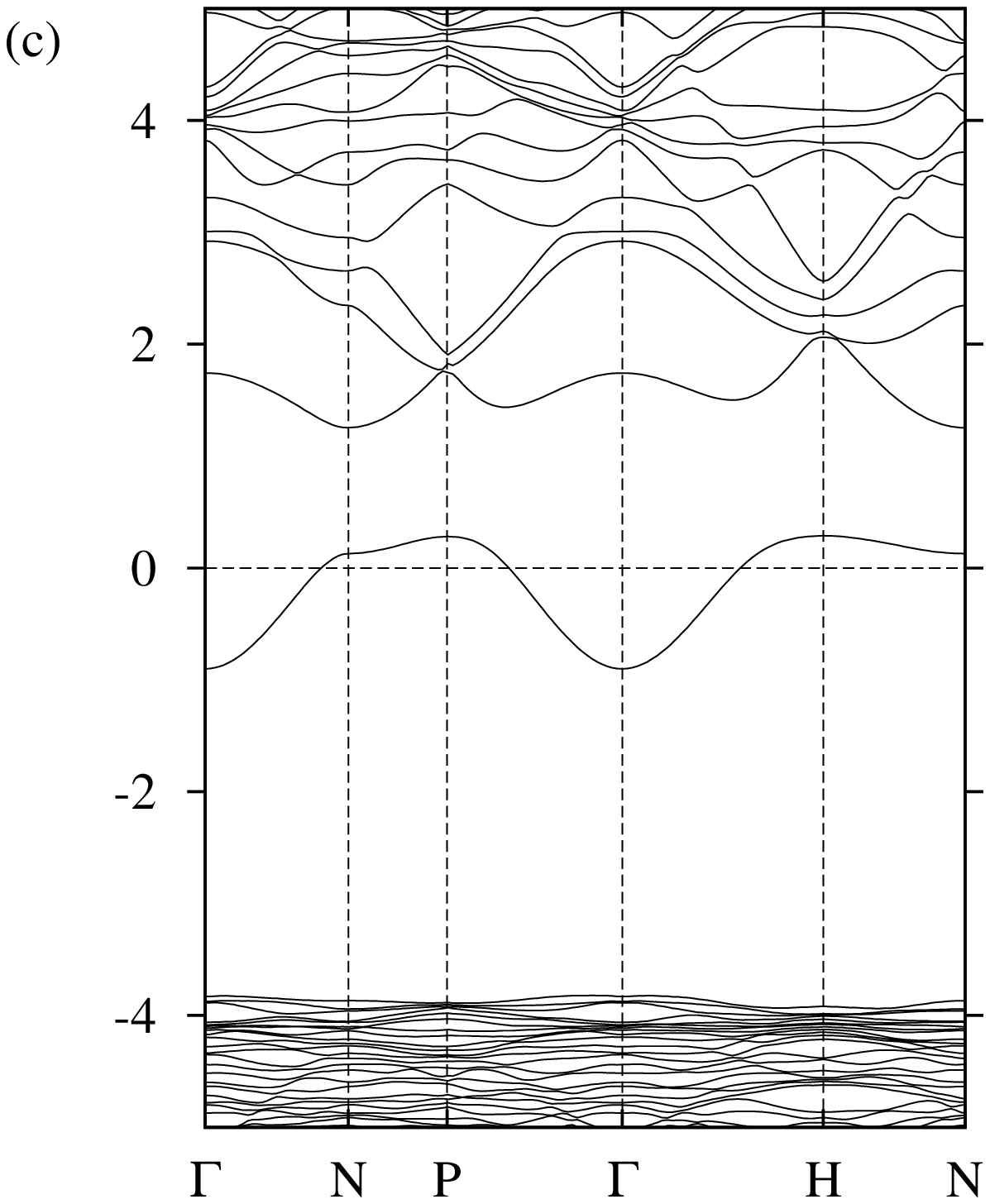}
\caption{Electronic band structure of (a) undoped stoichiometric In$_2$O$_3$,
(b) 6.25 at.\% Sn-doped In$_2$O$_3$ which corresponds to Sn
concentration of 1.93$\times$10$^{21}$cm$^{-3}$, 
and (c) oxygen deficient In$_2$O$_3$ with the oxygen 
vacancy V$_O^{+}$ concentration of 1.96$\times$10$^{21}$cm$^{-3}$
as obtained within self-consistent sX-LDA.}
\label{ino-sn-vac}
\end{figure*}

At the end of this section, we would like to point out the differences 
in the electronic band structures of oxygen deficient and substitutionally 
doped oxides.
Figure \ref{ino-sn-vac} illustrates how the electronic band structure of 
the In$_2$O$_3$ host is affected by the presence of oxygen vacancy 
(V$_O^{+}$) or substitutional Sn atom (Sn$_{In}$). 
For the same concentration and the same charge state of the two defects, 
we find that oxygen vacancy leads to a non-rigid-band shift of the Fermi 
level, and a large second gap appears due to the splitting 
of the lowest single conduction band 
from the rest of the conduction band, Fig. \ref{ino-sn-vac}. 
Conversely, there is no second gap in the case of Sn substitution.
We have found earlier \cite{my2prl} that the second gap previously 
reported for Sn-doped In$_2$O$_3$ \cite{Mryasov} 
vanishes upon structural relaxation around Sn ions. 
In marked contrast to the substitutional doping, 
the second gap {\it increases} by about 23 \% when the structure of 
the oxygen deficient In$_2$O$_3$ is allowed to fully relax.
Indeed, the absence of a second gap in case of the substitutional doping
indicates a good hybridization of the dopant states (Sn $s$-states) with 
the states of the neighboring atoms (O $p$-states), 
while an oxygen vacancy implies a substantial redistribution of 
the electron density to accommodate the defect and to minimize its impact.
The obtained differences in the structure of the conduction band bottom of
the doped and oxygen deficient oxide will, clearly, give rise to different 
transport and optical properties.

\begin{table*}
\caption{
Electronic properties of oxygen deficient In$_2$O$_3$, wurtzite and rock-salt 
ZnO, $\alpha$-Ga$_2$O$_3$ and $\beta$-Ga$_2$O$_3$, and InGaZnO$_4$
as calculated within LDA. 
The conduction band width reflects the degree of electron localization.
The smaller the $s$-orbital contribution on the cations nearest to the oxygen 
vacancy, the more charge stays trapped near the defect. 
}
\label{table-vacancy}
\begin{center}
\begin{tabular}{l|ccc|cc|c} \hline
                                 & In$_2$O$_3$ &  $h$-ZnO  & $\beta$-Ga$_2$O$_3$ &  $c$-ZnO & $\alpha$-Ga$_2$O$_3$ & InGaZnO$_4$  \\ \hline
Oxygen defect concentration, $\times10^{21}cm^{-3}$ 
                                  &   0.97   &  1.00 &    0.80    &  0.80  & 1.30 &  1.11 \\
Distance between the defects, \AA
                                  &    10    &   8   &      9     &   12   &  11  &  11 \\
Location of non-conducting V$_O^0$ state below &&&&&& \\
CBM, eV
                                  & 0.57 & 0.82 & 1.81 & 0.47 & 1.82 & 0.47 \\
Optical exitation V$_O^0 \rightarrow V_O^{+}+e^-$, eV	       
                                  &   1.64   &  1.48 &    2.01    &  1.68  & 2.19 &  1.64 \\ 
Location of metastable conducting V$_O^{+}$ state & & & & & & \\
below CBM, eV 
                                  & 0.38 & 0.35 & 1.39 & 0.49 & 0.88 & 0.31 \\ \hline
{\it For conducting oxides, i.e., those with} V$_O^{+}$ {\it defect}: &&&&&& \\
Conduction band width, eV         &   1.56   &  0.96 &    0.45    &  1.16  & 0.65 &  1.14 \\      
Electron velocity at Fermi level, $v_{BZ}\times 10^5 m/s$  
                                  &   6.67   &  3.91 &    0.96    &  6.47  & 2.35 &  4.76 \\
Direction of $v_{BZ}$ in standard Brillouin zone                                                             
                                &[1$\bar{1}$1]&[100]&[$\bar{1}$10]& [110] & [110] & [110] \\ 


Charge build-up at defect neighbors, \%   
                                       &  31   &  39  &  61    &  27  & 53 &  37  \\      
Nearest cations $s$-orbital contribution, \%  
                                       &  81   &  71  &  56    &  82  & 66 &  82(In), 74(Zn) \\ \hline
\end{tabular}
\end{center}
\end{table*}

\subsubsection{III.3 Electron localization in oxygen deficient TCO}

Our goal is to understand the microscopic origin of the different 
electronic properties (cf. Table \ref{table-vacancy}) 
of the three oxides in which the cations 
have the same electronic configuration, i.e., $s^0p^0$.
In particular, we would like to understand the origin of the strong 
electron localization that manifests itself in the narrow conduction 
band in oxygen deficient $\beta$-Ga$_2$O$_3$. For this,  
we analyze the charge density distribution within the lowest conduction band 
and also compare the contributions to the conduction band 
from the cations and anions located at different 
distances from the oxygen defect. 

For In$_2$O$_3$ we find that the In atoms nearest to the vacancy site
give 2-3 times larger contributions than the rest of the In atoms 
in the cell. As a result, there is a notable build-up of the charge
density near the vacancy, namely, 31 \% of the total charge at the bottom 
of the conduction band reside on the nearest neighbor In and 
the next nearest neighbor O atoms.

In oxygen deficient ZnO, there is an order of magnitude difference between 
the contributions from the specific atoms (either cations or anions), 
and 39 \% of the total charge density belong to the nearest Zn and
next nearest O neighbors of the oxygen vacancy. 

In Ga$_2$O$_3$ we find that  61 \% of the total charge is  
at the defect's nearest neighbor cations  
and the next nearest neighbor oxygen atoms, 
with the rest of the cations or anions giving up to two orders of magnitude 
smaller contributions.

Thus, the electron density distribution is more uniform within the cell 
in indium and zinc oxides, whereas a strong electron localization 
is observed in the oxygen deficient Ga$_2$O$_3$. For further analysis, 
we consider the role of the crystal symmetry (Section III.4) 
and the type of orbitals ($s$, $p$ or $d$) which form the conduction band
(Section III.5).

\subsubsection{III.4 Role of the crystal symmetry}

The trend in the amount of the build-up charge near the vacancy site
and the resulting energy dispersion of the conduction band and associated 
electron group velocities (Table \ref{table-vacancy}) in 
the single-cation oxides correlate with their crystal symmetry, i.e.,  
cubic In$_2$O$_3$ $>$ hexagonal ZnO $>$ monoclinic Ga$_2$O$_3$. A higher symmetry 
is expected to provide a larger orbital overlap and, hence, a better charge transport. 
We wish to shed some light onto this correlation.

\subsubsection{III.4.A Long-range symmetry: crystal anisotropy}

First, let us consider the role of the long-range crystal lattice anisotropy 
which is exists in $\beta$-Ga$_2$O$_3$ (monoclinic phase, space group $C2/m$)
due to the presence of structurally distinct (non-equivalent) Ga and O atoms.
As one can see from Table \ref{table-hosts}, the effective mass anisotropy
is the largest for the undoped stoichiometric $\beta$-Ga$_2$O$_3$, with
the masses found to be equal to m$^{[100]}$=m$^{[010]}$=0.35 m$_e$ 
and m$^{[001]}$=0.32 m$_e$ within sx-LDA. 
However, this effective mass difference in the oxide host cannot explain 
the obtained strong anisotropy of the electronic properties that appears 
in the oxygen deficient $\beta$-Ga$_2$O$_3$ as discussed below.

Due to the presence of the non-equivalent Ga and O atoms, 
the distribution of the oxygen vacancies is not uniform. 
We find that V$_O^{+}$ prefers to be located in the oxygen sites of the O(1) type,
while the total energy of the configuration with the defect in the O(2) or O(3) sites is 
higher by 163 meV or 267 meV, respectively.
The oxygen atoms of the O(1) type form zig-zag chains along the $y$ direction with the atomic
distance within the chain of 3.04 \AA, while the shortest distances between the chains 
are 3.91 \AA \, and 4.76 \AA \, in the $x$ and $z$ directions, respectively.
Therefore, the interaction between the oxygen vacancies distributed among the O(1) sites 
should be stronger within the chains than between the chains. This leads to 
different Fermi electron group velocities along the main crystallographic 
directions: we obtain  
0.8$\times$10$^5$ $m/s$, 1.9$\times$10$^5$ $m/s$ and 1.3$\times$10$^5$ $m/s$ 
along the $x$, $y$ and $z$ directions, respectively. 
The conductivity which is proportional to the square of the electron velocity 
is expected to be highly anisotropic -- with the factor of 6 between 
the $x$ and $y$ directions.

Thus, we argue that crystal lattice anisotropy affects the transport properties 
of a TCO primarily due to a non-uniform distribution of carrier donors 
whereas the electronic properties of the TCO host remain nearly isotropic.
For each compound, there may be a particular carrier generation mechanism 
found which maintains isotropic nature of the oxide host.

\subsubsection{III.4.B Local symmetry: oxygen coordination}

\begin{figure}
\includegraphics[width=5.3cm,angle=-90]{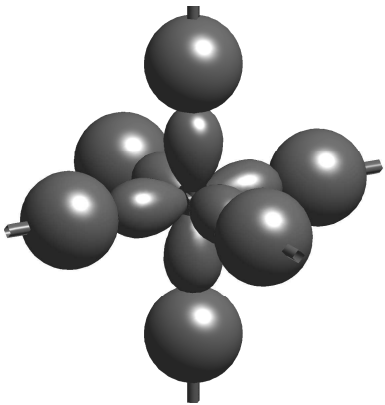}
\caption{Octahedral coordination of oxygen atom 
by cations provides the largest overlap between the $s$-orbitals of 
the six metal ions and $p_x$, $p_y$ and $p_z$ orbitals of the oxygen atom.}
\label{overlap}
\end{figure}

As discussed in Section III.1, the conduction band in the oxides 
under investigation is formed from the $s$ orbitals of the cation 
and $p$ orbitals of the oxygen atoms. These orbitals will participate
in the charge transport once the materials are degenerately doped. 
The orbital overlap between the neighboring atoms 
depends on the oxygen coordination. From the symmetry considerations, 
an octahedral coordination provides better $s$-$p$ overlap, 
Figure \ref{overlap}, compared to a tetrahedral one. 
Therefore, the energy dispersion of the conduction band is expected 
to be larger in the case of octahedral coordination.  

To elucidate the role of oxygen coordination in the charge 
transport properties of the oxides, we performed calculations 
for the oxygen deficient high-pressure $\alpha$-Ga$_2$O$_3$ phase 
with rhombohedral Al$_2$O$_3$-type structure (space group $R\bar{3}c$)
and rock-salt ZnO phase. (In$_2$O$_3$ with $R\bar{3}c$ 
(corundum) and $I2_13$ structures were not considered here because 
the cations (anions) have the same coordination, i.e., octahedral 
(tetrahedral), as in bixbyite structure.)

In high-symmetry $\alpha$-Ga$_2$O$_3$ phase 
all Ga atoms have octahedral oxygen coordination -- 
in contrast to $\beta$-Ga$_2$O$_3$ with monoclinic $C2/m$ structure which possesses
both tetragonally and octahedrally coordinated Ga atoms.
We find that in the oxygen deficient $\alpha$-Ga$_2$O$_3$ with
V$_O^{+}$ concentration of 1.3$\times$10$^{21}$cm$^{-3}$, 
the nearest Ga and O atoms trap 53 \% of the total charge density 
which is smaller compared to 61 \% in $\beta$-Ga$_2$O$_3$ 
with similar defect concentration of 0.8$\times$10$^{21}$cm$^{-3}$. 
Consequently, the width of the single conduction band increases from 0.45 eV
in $\beta$-Ga$_2$O$_3$ to 0.65 eV in $\alpha$-Ga$_2$O$_3$ resulting in more than twice 
larger electron group velocity for the high-symmetry phase, 
Table \ref{table-vacancy}.

Similarly, we find that in oxygen deficient cubic ZnO with octahedral oxygen 
coordination of cations, the contributions from the atoms nearest 
to the defect decrease to 27 \% of the total charge density and 
the contributions from atoms located at different distances from 
the oxygen defect differ by only 2 or 3 times.
Accordingly, the conduction band width is larger (1.16 eV) that
leads to a larger electron velocity near the Fermi level as compared to 
the ground-state hexagonal ZnO with the tetragonal oxygen coordination 
around Zn atoms (Table \ref{table-vacancy}).

Thus, we conclude that the high symmetry of octahedral oxygen coordination
may improve conductivity owing to a better $s$-$p$ overlap which leads 
to a wider conduction band and, hence, a larger electron velocity. 
A way to utilize the potential of an oxide is to ensure that the cations and anions possess a high {\it local} symmetry which can be attained  
via stabilization of a higher-symmetry phase or by choosing a multicomponent material 
with octahedral oxygen coordination for all constituents.

\begin{figure*}
\includegraphics[height=5.0cm]{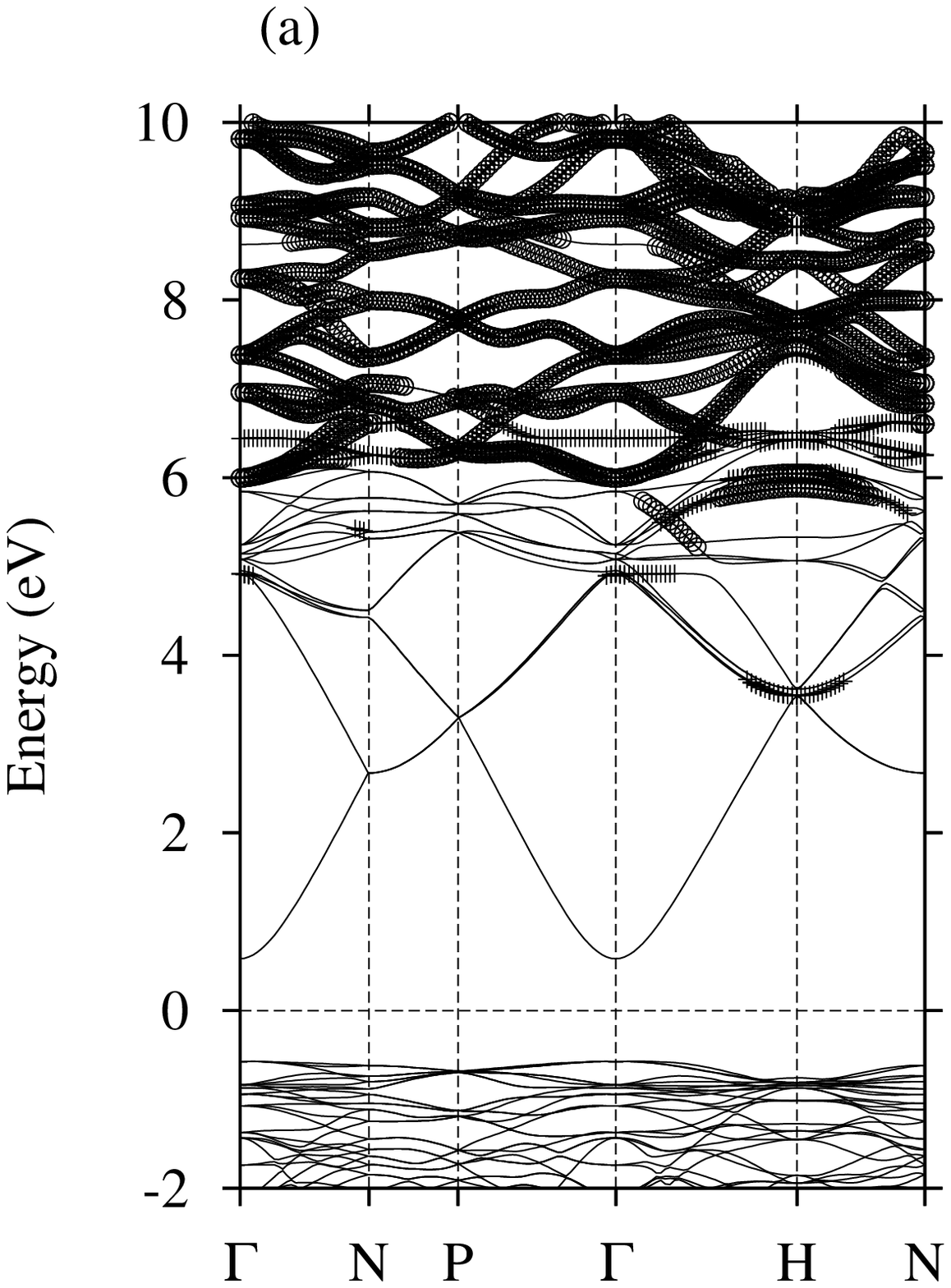}
\includegraphics[height=5.0cm]{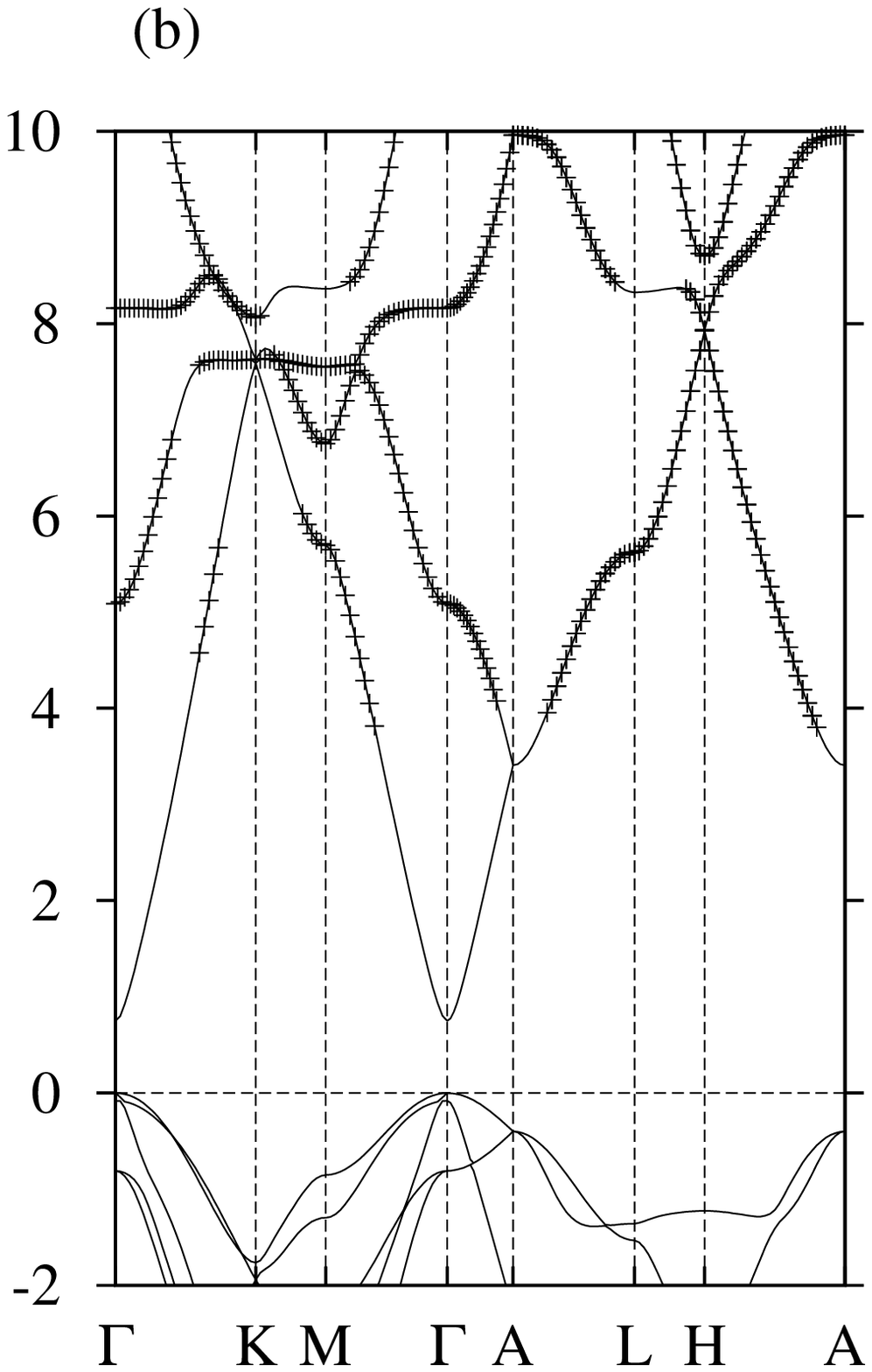}
\includegraphics[height=5.0cm]{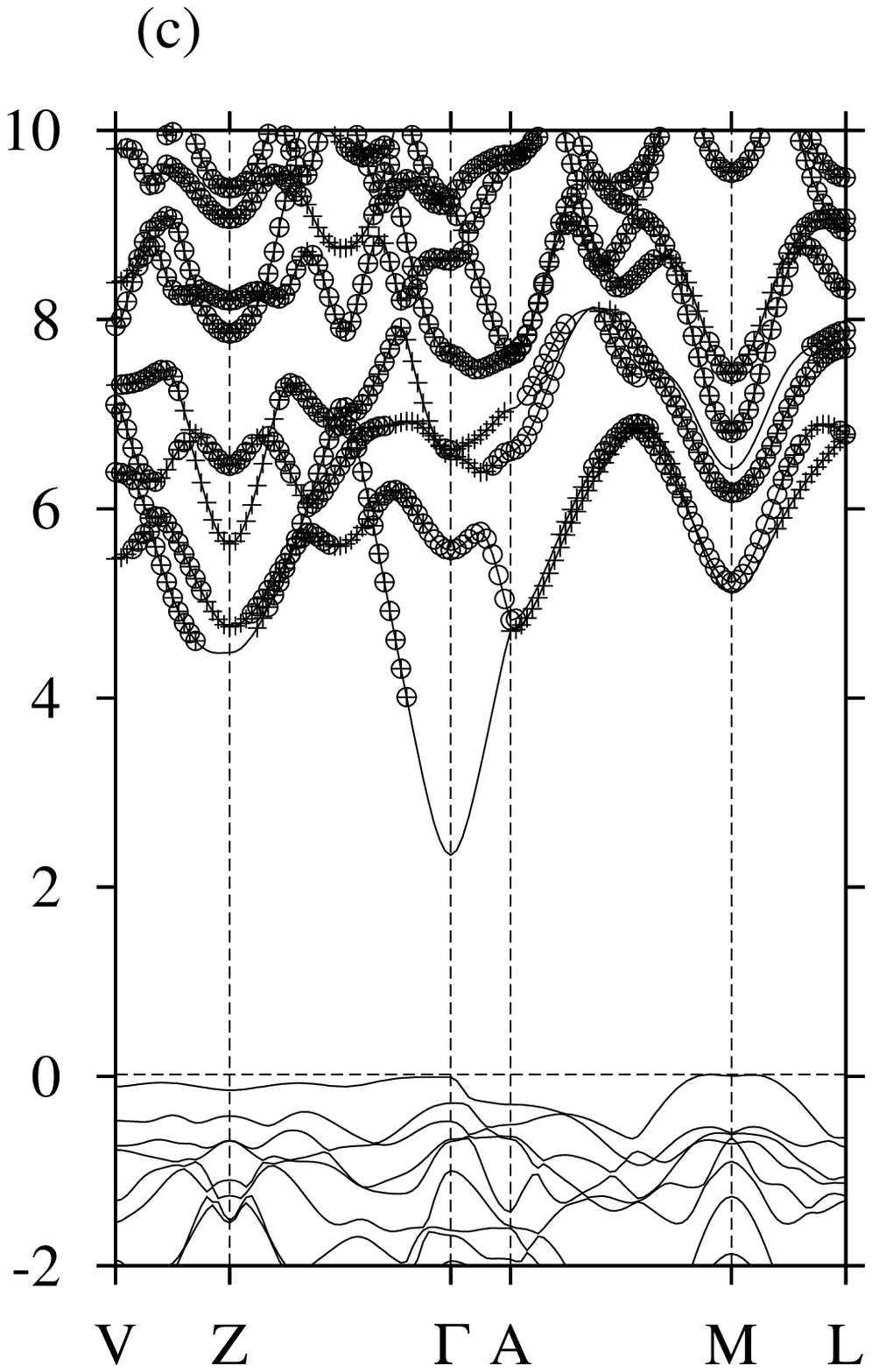}
\includegraphics[height=5.0cm]{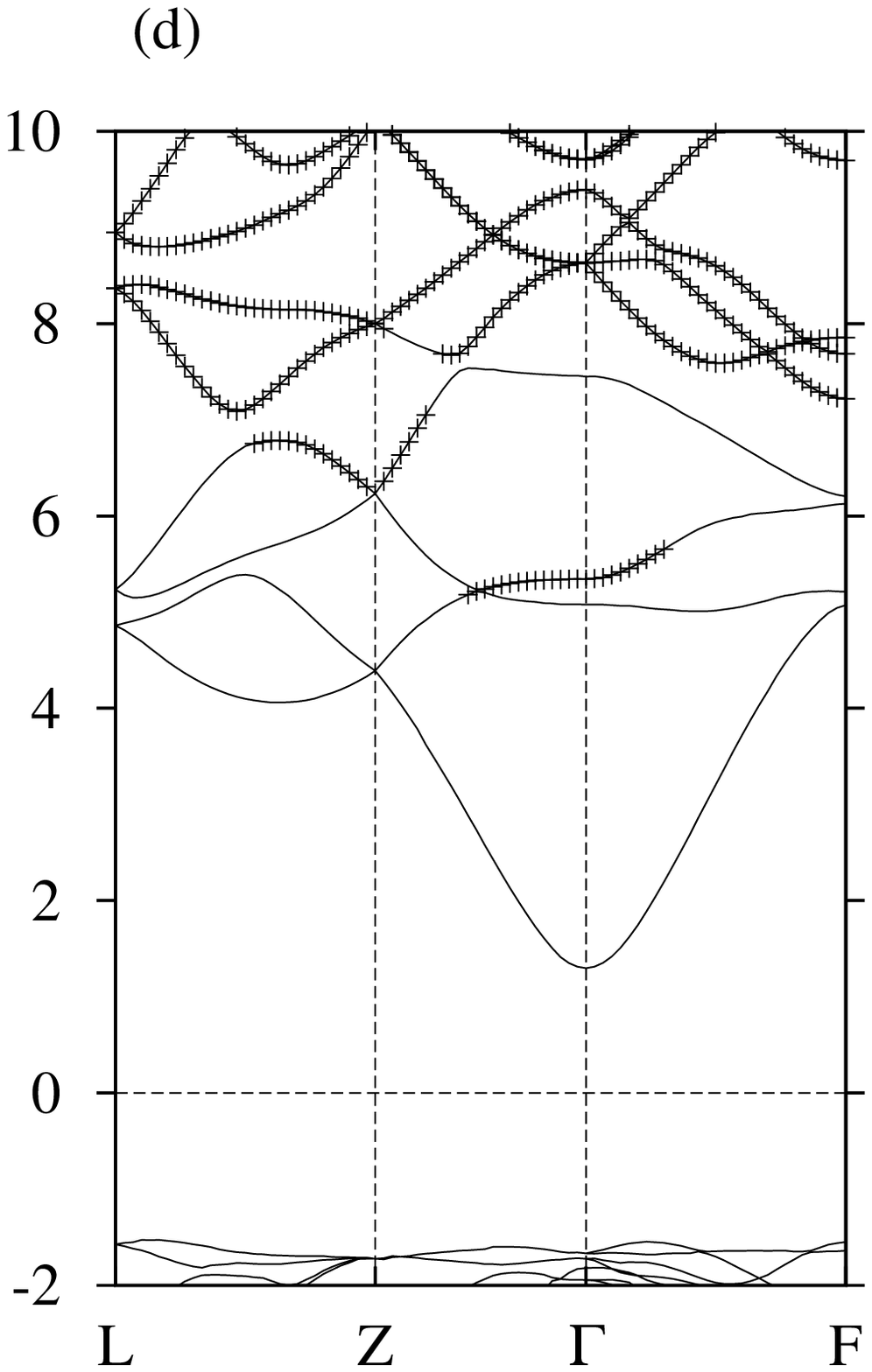}
\includegraphics[height=5.0cm]{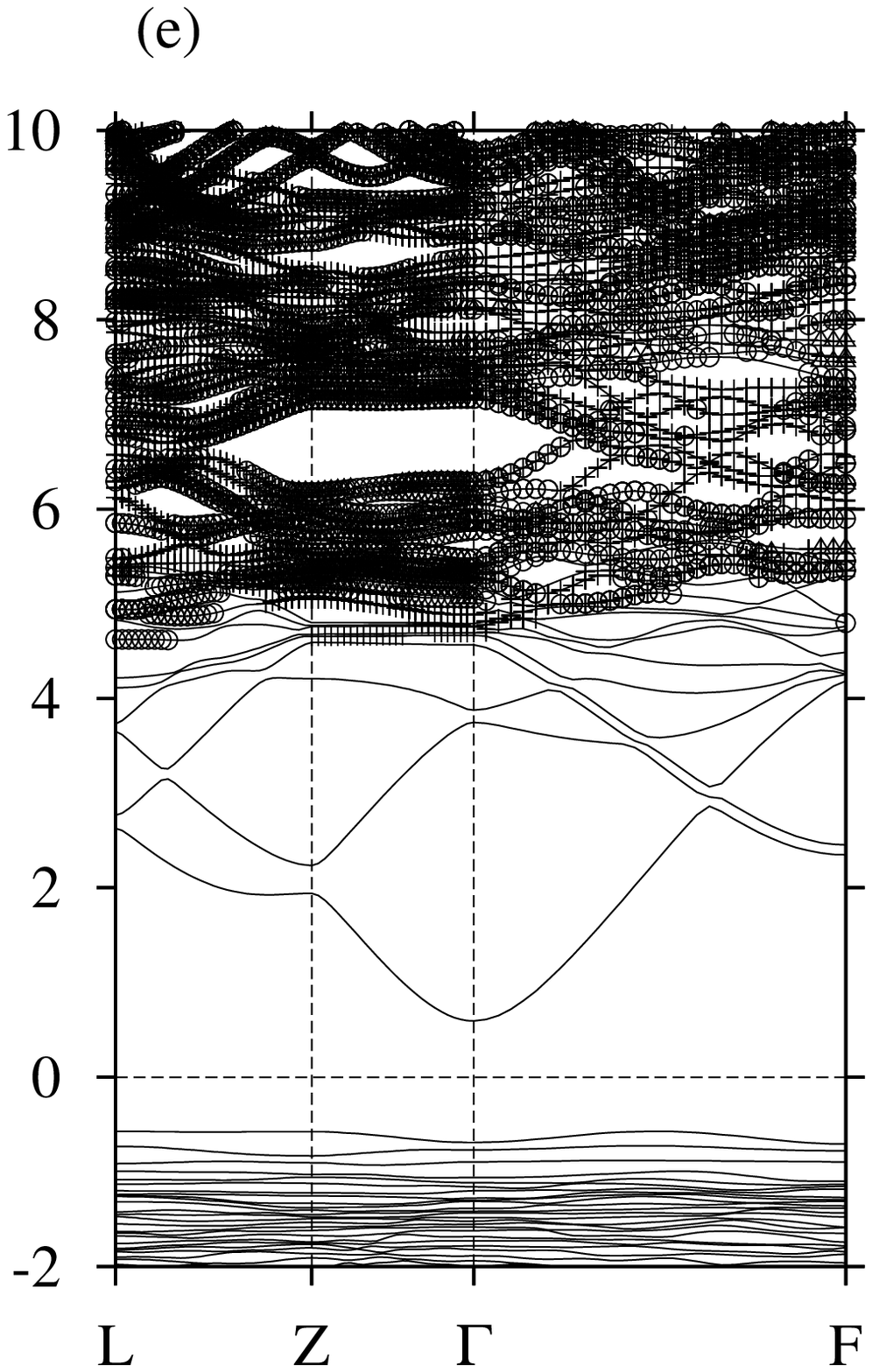}
\caption{Band structure of undoped stoichiometric (a) In$_2$O$_3$, 
(b) wurtzite ZnO, (c) $\beta$-Ga$_2$O$_3$, (d) $\alpha$-Ga$_2$O$_3$,
and (e) InGaZnO$_4$ as calculated within LDA. 
The empty $p$-states of cation(s) are marked 
with ($+$), ($\bigcirc$) or ($\bigtriangleup$) symbols to differentiate 
between the structurally and/or chemically distinct cations.
Remarkably, 
the hybrid conduction band in InGaZnO$_4$ is formed well below 
the detrimental Ga $p$-states giving rise to a good charge transport.
}
\label{p-states}
\end{figure*}

\subsubsection{III.5 Role of cation's orbital contributions}

Despite the octahedral coordination of the cations in rhombohedral 
Ga$_2$O$_3$, the electron velocity in the oxygen deficient material
still differ by almost a factor of 3 from those in In$_2$O$_3$ 
and rock salt ZnO, Table \ref{table-vacancy}.
Again, the electron effective masses of stoichiometric oxides, 
cf. Table \ref{table-hosts}, cannot explain the differences.
We extend our investigation by comparing the cation's orbital
contributions to the conduction band.

As mentioned in Section III.1, the cations and anions give nearly the same contributions  
at the bottom of the conduction band in all oxides investigated. 
Now we consider oxygen deficient materials and analyze the orbital 
contributions from cations, in particular, those that are nearest to the oxygen vacancy. (The contributions from these atoms are considerable, judging by the charge build-up at the defect neighboring atoms which is observed for all oxides under investigation, see Section III.3 and Table \ref{table-vacancy}).
Most importantly, we find that not only the $s$ states but also the $p$ and $d$ states of 
the cations contribute to the bottom of the conduction band near the $\Gamma$ point. 
This is expected from the {\bf k$\cdot$p} theory, see for example \cite{kp}, which suggests that 
the states located both higher (conduction states) and lower (valence states) in energy with respect 
to a specified band (the conduction band minimum in our case) 
may contribute to that band. (Note that the sum in Eq. (1) runs over 
{\it all} bands except the conduction band).

We compare the orbital contributions from the cations to the conduction band wave functions
in the oxygen deficient In$_2$O$_3$, rock salt ZnO and rhombohedral Ga$_2$O$_3$.
For the latter two oxides we use the high-symmetry phases where the cations  
have octahedral coordination. This will allow us to separate the role of 
local symmetry (the oxygen coordination) from the role of the electronic 
configuration of the cations in the observed differences in the transport 
properties. Values for the ground-state wurtzite ZnO and $\beta$-Ga$_2$O$_3$ 
are also given in Table \ref{table-vacancy} for clarity.

For In$_2$O$_3$ and rock salt ZnO we find that more than 80\% of the nearest 
cation contributions comes from their $s$-orbitals. 
Therefore, the conduction band originates primarily from the hybridization 
between the $s$-states of the cation and the antibonding $p$-states of the oxygen atoms. 
Clearly, the $sp$ hybridization provides the most uniform charge distribution within the cell 
and, thus, facilitates good carrier transport.

In marked contrast to In$_2$O$_3$ and rock salt ZnO, in rhombohedral Ga$_2$O$_3$ 
the $s$-orbitals of the cations located near the oxygen defect contribute 
only 66\%, while the contributions from the Ga $p$-states become considerable, 
21 \%of the cations' total.
This means that a substantial part of the charge introduced via the oxygen vacancy 
becomes trapped on the $p$ orbitals of the neighboring Ga atoms. 
The charge localization occurs due to the weak hybridization (small overlap) of these highly anisotropic $p$-orbitals with the $p$-orbitals of the oxygen atoms from the next coordination sphere.
Low crystal symmetry reduces this hybridization even further, as demonstrated by the results for $\beta$-Ga$_2$O$_3$ and wurtzite ZnO, Table \ref{table-vacancy}.


The origin of the suppressed $s$-orbital contributions in Ga$_2$O$_3$ and the resulting electron trapping near the defect is the proximity of the Ga $p$ states to the conduction band bottom that gives rise to larger contributions from these $p$-orbitals.
In Fig. \ref{p-states}, we compare the calculated electronic band structures for undoped stoichiometric oxides where the cation's empty $p$-states are highlighted. We find that the $p$-states in Ga$_2$O$_3$ are located more than 1 eV closer to the bottom of the conduction band as compared to In$_2$O$_3$ and ZnO.
Therefore, in contrast to indium and zinc oxides, the cation $p$-states in Ga$_2$O$_3$ may become 
available for the extra electrons -- as we, indeed, found.

Based on the above findings for single-cation TCOs, we can conclude that 

(i) the hybridization between the cation's $s$-orbitals and the $p$-orbitals
of the neighboring oxygen atoms is crucial for achieving good charge transport; 

(ii) when the anisotropic $p$ (or $d$) orbitals of the cation contribute 
considerably to the conduction band, the probability for the electron(s) 
to propagate further throughout the cell is reduced -- owing to a weak 
hybridization between these $p$ orbitals and the $p$ orbitals of the 
neighboring oxygen atoms. As a result, the electron velocity and, hence, 
conductivity are reduced. The degree of reduction depends on how close 
in energy the cation's $p$ (or $d$) states are located with respect 
to the bottom of the conduction band.

These findings explain the experimental observations that the conductivity 
of a multicomponent TCO decreases with increasing Ga content.

\subsection{IV. Electronic properties of InGaZnO$_4$}

\subsubsection{IV.1 Undoped stoichiometric InGaZnO$_4$}

Electronic band structure investigations of undoped stoichiometric 
InGaZnO$_4$ have been performed
earlier for the crystalline \cite{my2epl,Hosono-defects-vasp} and 
amorphous oxide \cite{Kamiya-Hosono-amorph}. Most importantly, it was found 
that despite the different band gap values of the constituent single-cation 
oxides, cf., Table \ref{table-hosts}, all cations give comparable 
contributions to the conduction band bottom of the complex oxide. 
Because LDA may fail in description of the empty conduction states, 
the above finding of the hybrid nature of the conduction band should 
be verified within sX-LDA.
Here we performed such calculations for InGaZnO$_4$ and found that although 
the contributions (to the conduction band at $\Gamma$ point) from 
In states increase -- from 23 \% in LDA to 33\% in sX-LDA, those from 
Ga and Zn remain the same (15-16 \% and 11 \%, respectively) and so 
these states will be available for extra electrons once those are introduced
into the host oxide. 
The change in the In contributions arises due to reduced contributions 
from O(1) and O(2) -- namely, from 26 \% in LDA to 23 \% in sX-LDA and 
from 24 \% in LDA to 17 \% in sX-LDA, respectively. The reduction in 
the oxygen contributions is expected from a larger band gap in sX-LDA 
which we obtained to be equal to 3.29 eV, in good agreement with the 
experimental value of 3.5 eV \cite{Orita-exp}. 
Finally, the electron effective mass remains isotropic in sX-LDA, 
Table \ref{table-hosts}, confirming that both structurally and chemically 
distinct layers, InO$_{1.5}$ and GaZnO$_{2.5}$, will participate 
in charge transport upon a degenerate doping of this material.

\subsubsection{IV.2 Spatial distribution of oxygen defect in InGaZnO$_4$}

\begin{figure}
\includegraphics[width=8cm]{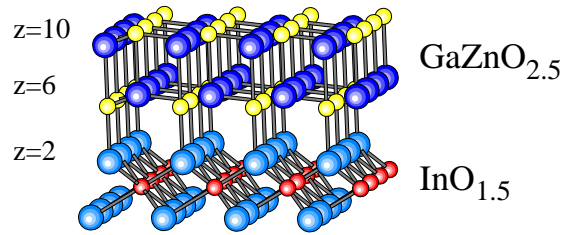}
\caption{
Crystal structure of InGaZnO$_4$. Only one of the three blocks 
which construct the conventional unit cell when stacked along 
the $z$ direction, is shown. The cartesian $z$ coordinates 
of the oxygen atoms are given on the left.
}
\label{str}
\end{figure}

First, we investigate the spatial distribution of the oxygen vacancy in the complex 
layered InGaZnO$_4$ by determining the most energetically favorable location of 
the oxygen vacancy in the lattice. There are four structurally different sites 
for the defect, Fig. \ref{str} and Table \ref{table-vacancy-complex}.
It can be located within the InO$_{1.5}$ layer (this corresponds to a vacancy in 
the O(1) structural type \cite{str}) or GaZnO$_{2.5}$ layers (a vacancy in the O(2) site). 
In addition, because Ga and Zn atoms are distributed randomly in the GaZnO$_{2.5}$ layers \cite{Li},
oxygen vacancy can have either Ga$^{3+}$ or Zn$^{2+}$ as an apical cation, i.e., the cation 
in the layer adjacent to the one where the defect is located. 
A comparison of the total energies of the oxygen deficient systems reveals that
for all three charge states of V$_O$, the defect prefers to be in the InO$_{1.5}$ layer
having Zn as its apical atom, Table \ref{table-vacancy-complex}.
We note that the energy difference between the oxygen vacancy located in InO$_{1.5}$ 
and the one in GaZnO$_{2.5}$ layers 
increases significantly as the defect charge state decreases: it is only 56 meV for uncompensated
V$_O^0$, while for partially or completely compensated defect the difference 
is an order of magnitude larger, Table \ref{table-vacancy-complex}.

\begin{table}
\caption{Respective total energy, in meV, of InGaZnO$_4$ system 
with uncompensated, partially compensated and fully compensated oxygen vacancies 
located at structurally different sites.
}
\label{table-vacancy-complex}
\begin{center}
\begin{tabular}{lrrr} \hline
Defect location & V$_O^0$ &  V$_O^{+}$ & V$_O^{2+}$ \\ \hline

Within the InO$_{1.5}$ layer     & & & \\
\,\,\, apical Ga             &     +194      &     +138       &   +267 \\
\,\,\, apical Zn             &        0      &        0       &      0 \\   

Within the GaZnO$_{2.5}$ layers   & & & \\
\,\,\, apical Ga             &     +110      &     +793       & +1,319 \\
\,\,\, apical Zn             &      +56      &     +536       &   +730 \\ \hline
\end{tabular}
\end{center}
\end{table}

An important observation is that the energetics of different site 
locations of the oxygen defect in InGaZnO$_4$ correlates with the formation 
energy of the constituting single-cation oxides, namely, 
--6.42 eV/O$_2$ for In$_2$O$_3$, --7.20 eV/O$_2$ for ZnO and --7.46 eV/O$_2$ 
for Ga$_2$O$_3$ \cite{Reed}.
The most preferred location of the vacancy (i.e., the O(1) site with three In and one Zn neighbors) 
corresponds to the selection of the metal-oxygen bonds that would be easiest to break in order to create the defect. 
Although this conclusion must be verified for other systems, it may serve as initial guidance in defect studies of complex TCOs, both in crystalline and amorphous forms.


From the formation energy calculations we find that the uncompensated oxygen 
vacancy, V$_O^0$, corresponds to the ground state of 
the defect -- similar to the In$_2$O$_3$, ZnO and Ga$_2$O$_3$ cases, 
Fig. \ref{formation}. 
The LDA-derived energy required to excite V$_O^0$ to obtain 
V$_O^{+}$ as well as the location of the oxygen defects with respect to
the conduction band bottom in InGaZnO$_4$ are also similar to the single-cation oxides, 
Table \ref{table-vacancy}. Hence, the vacancy can be photoexcited to provide 
a metastable conductive state.
Investigations of possible charge-compensating native defects 
(e.g., V$_{In,Ga,Zn}$ or Zn$_{Ga}$) and  
defect complexes (e.g., (Zn$_{Ga}$V$_O$)) 
are required to further understand the lower observed conductivity 
in the layered InGaZnO$_4$ and in similar compounds from the series
as compared to the single-cation TCOs.

Finally, the obtained preferential distribution of the oxygen vacancies in InGaZnO$_4$ may explain the observed strong anisotropy of the electrical 
conductivity for the layered In$_2$O$_3$(ZnO)$_n$ compounds with an increased 
number of ZnO layers, $n\geq$3 \cite{thermoelectr,WeiPRB,Hiramatsu}.
A larger spatial separation between the InO$_{1.5}$ layers due to extra ZnO layers
may reduce the interaction between the oxygen defects which prefer to reside
in the InO$_{1.5}$ layers. This will lead to a smaller electron velocity along the $z$-direction.
We note, however, that the anisotropy of the transport properties may not be well-defined in the as-grown samples since oxygen vacancy may occupy other, less favorable, locations. 
A higher but more anisotropic conductivity can be attained via a post-growth annealing with temperatures sufficient for the defect migration. Indeed, the observed significant decrease in the electrical 
conductivity along the $z$ axis with temperature which is reported to be almost twice smaller 
than the conductivity within the $ab$ plane in the high temperature region \cite{thermoelectr} 
supports our conclusion.

\subsubsection{IV.3 Electronic properties of oxygen deficient InGaZnO$_4$}

Now we analyze the conduction band wave function 
in the oxygen deficient InGaZnO$_4$ -- similar to the studies 
for single-cation oxides, see Section III.  
In accord with the hybrid nature of the bottom of the conduction band   
in the undoped stoichiometric InGaZnO$_4$, see \cite{my2epl} and Section IV.1, 
all atom types in the cell give comparable contributions to the conduction 
band wave function: 
In, Ga, Zn, O(1) and O(2) atoms contribute 25 \%, 11 \%, 10 \%, 32 \% and 
22 \%, respectively. 
Moreover, for each type of atoms the contributions are fairly similar: 
the largest deviations are a factor of 2.5, 1.5, 2.0, 4.6 or 1.6 for, 
respectively, In, Ga, Zn, O(1) or O(2) located at different distances 
from the oxygen defect. Therefore, the charge density in the conduction 
band of the oxygen deficient InGaZnO$_4$ is uniformly distributed throughout 
the cell -- similar to In$_2$O$_3$ and ZnO cases and  
in contrast to $\beta$-Ga$_2$O$_3$ where the defect's nearest and next nearest 
neighbor atoms give up to two orders of magnitude smaller contributions than the rest of the cations and anions in the cell.
Accordingly, for multicomponent oxide we find that the energy dispersion of 
the conductive band and the electron velocities near the Fermi level 
are comparable to those obtained for oxygen deficient In$_2$O$_3$ and 
superior to those for wurtzite ZnO, Table \ref{table-vacancy}.

For InGaZnO$_4$ we find that the charge build-up at the defect's nearest 
cations and next nearest anions is 37 \% which is notably lower than 
in Ga$_2$O$_3$ in either $\alpha$ or $\beta$ phase, cf., 
Table \ref{table-vacancy}. It is important to point out that similar 
percentage is obtained for all locations of the oxygen defect, 
including the one where the vacancy has three Ga neighbors. 

Further, we analyze the relative orbital contributions from the defect neighboring cations to the conduction band wave function.  
We find that the $s$-orbitals of the In and Zn atoms nearest to the vacancy located in the InO$_{1.5}$ layer or of the Ga and Zn atoms nearest to the vacancy in the GaZnO$_{2.5}$ layer contribute
82\% and 74\% or 75\% and 71\% of the cations total, respectively. 
The considerable increase in the $s$-contributions from the Ga atoms -- as compared to the Ga$_2$O$_3$ phases, Table \ref{table-vacancy} --
is attributed to the hybrid nature of the bottom of the conduction band in InGaZnO$_4$. 
Due to the strong hybridization between the states of {\it every} cation with the states 
of its neighboring oxygen atoms and {\it because of the differences in the band gaps}
of the constituent oxides (3.1-3.6 eV for ZnO, 3.5-3.7 eV for In$_2$O$_3$ and 4.4-4.9 eV 
for Ga$_2$O$_3$, found experimentally), the hybrid band is formed well below the Ga $p$-states, 
cf., Fig. \ref{p-states}(d). Qualitatively, this means that the interaction
between the Ga $s$-states (conduction band) and the O $p$-states (valence band) 
become weaker in InGaZnO$_4$ (band gap 3.3 eV) as compared to Ga$_2$O$_3$ 
(band gap $\sim$4.6 eV), 
while the cation's $p$-states remain located at about the same energy, Fig. \ref{p-states}. 
As a result, the contributions to the conduction band wavefunction from 
these detrimental Ga $p$ orbitals are notably smaller in InGaZnO$_4$ 
(14\%) as compared to Ga$_2$O$_3$ (33\% or 21\% in $\beta$ or $\alpha$ phase, respectively).
Thus, the conduction band in complex InGaZnO$_4$ consists primarily from 
the cations $s$ and the oxygen $p$ states. This explains the obtained large electron velocity 
in this material, Table \ref{table-vacancy}.

The above results highlight an important advantage of multicomponent oxides: 
mixing several oxides with notably different band gaps may help overcome the charge 
localization effects by tuning the resulting electronic properties via chemical composition
of the complex material (and, to a lesser extent, via choosing/stabilizing a proper crystal structure).
We must stress that both the relative content of the constituent oxides as well as their 
band gaps are critical parameters which control the resulting optical and transport properties. 
A large content of an oxide with a wider band gap compared to that of the other constituents 
may be appealing for the optical properties as well as for optimization of the material's work function
desired for practical applications \cite{Inoue}. 
At the same time, however, as the band gap increases, the bottom of 
the hybrid conduction band will be effectively moved up -- closer to 
the cation(s) $p$-states. 
This will increase the $p$-orbital contribution to the conduction band 
and limit the carrier mobility due to the increased probability for the 
electrons to be trapped on the cation's $p$-states.

\subsection{V. Substitutional doping in InGaZnO$_4$}

Targeted doping with specific impurities 
which donate extra electrons upon aliovalent substitution represents 
an alternative route for the carrier generation in TCO hosts.
Compared to the native point defects such as oxygen vacancy,
substitutional doping is more appealing due to better stability 
of the doped samples to the environment that is essential for practical applications. 
Among single-cation oxides, the most studied and commercially utilized 
are Sn-doped In$_2$O$_3$, Al-doped ZnO, Sb-doped SnO$_2$, F-doped ZnO and SnO$_2$. 
To the best of our knowledge, there were no reports
on substitutional doping of InGaZnO$_4$ or other materials from the homologous series
(in Ref. \cite{thermoelectr}, (ZnO)$_m$In$_2$O$_3$ was doped with 
Ca$^{2+}$, Sr$^{2+}$ and Ba$^{2+}$ to improve the thermoelectric properties). 

One of the main challenges of substitutional doping in a multicomponent oxide 
is to find a proper dopant which will donate
extra electron(s) and lead to degenerate doping. The valence and the ionic size
of both the dopant and all the cations in the cell should be taken into consideration.
For example, doping of InGaZnO$_4$ with Al, which is known to be a good choice in the case of ZnO, 
is likely to result in Al$^{3+}$ (the ionic radius is 0.53 \AA) substituting Ga$^{3+}$ (0.61 \AA) 
instead of Zn$^{2+}$ (0.74 \AA) so that the extra electrons are not generated and 
the material remains non-conducting.
Moreover, non-uniform spatial distribution of the dopants, 
their clustering or secondary phase formation as well as formation of defect complexes 
may become more pronounced in multicomponent 
oxides due to complexity and peculiarities in their crystal structure
as compared to their single-cation counterparts.
Systematic theoretical investigations may prove to be
valuable to guide future experiments.

\begin{table}
\caption{Properties of subtitutionally doped InGaZnO$_4$.
The dopant ionic radius, R, in \AA; the optimized sistances between the dopant and its oxygen neighbors,
D(M-O), in \AA; the formation energy of the defect, E$_{form}$, in eV;
the electron velocities, $v$, calculated along 
the main crystallographic directions in the vicinity of the Fermi level; 
the density of states at the Fermi level, N(E$_F$), in states/eV$\cdot$cell;
the fundamental band gap, E$_{fundamental}$, in eV; the Burstein-Moss (BM) shift 
(the Fermi energy displacement measured from the bottom of the conduction band), in eV, 
for Ga, Sn, Ti and Zr doped InGaZnO$_4$.
The dopant concentration is 1.75$\times$10$^{21}$ $cm^{-3}$ 
in all materials. For comparison, the ionic radii of In and Zn are 
0.94 \AA \, and 0.74 \AA, respectively.}
\label{subst}
\begin{center}
\begin{tabular}{lcccc} \hline
                          & Ga$_{Zn}$ & Sn$_{In}$  &  Ti$_{In}$  &  Zr$_{In}$  \\ \hline
R  & 0.61 & 0.83 & 0.75 & 0.86 \\
D(M-O)  & 1.88--2.16 & 2.09--2.15  & 1.98--2.10 & 2.05--2.45 \\
E$_{form}$                   & --0.32 & 0.62 & --2.79 & --3.92 \\
$v^{[100]}\times 10^5 m/s$      & 8.56 & 8.18 & 4.93 & 7.89 \\
$v^{[010]}\times 10^5 m/s$      & 8.69 & 8.11 & 5.32 & 8.17 \\
$v^{[001]}\times 10^5 m/s$      & 8.75 & 8.52 & 5.79 & 8.12 \\
N(E$_F$)      & 2.2 & 2.4 & 6.3 & 2.3 \\
N(E$_F$)$v^2$ & 3.63 & 3.65 & 3.97 & 3.31 \\
E$_{fundamental}$ & 1.26 & 1.20 & 1.43 & 1.48 \\
BM shift, eV   & 1.51 & 1.45 & 1.07 & 1.41 \\ \hline
\end{tabular}
\end{center}
\end{table}

Our goal is to investigate the structural, electronic and optical properties of doped InGaZnO$_4$
and also to compare them with those obtained for the oxygen deficient material.
We begin by studying Sn$^{4+}$ and Ga$^{3+}$ substitutions. 
The former dopant is the most intuitive choice, while 
extra Ga may be considered as an intrinsic stoichiometry defect. 
In addition to these ``traditional'' dopants, we also considered Ti$^{4+}$ 
and Zr$^{4+}$ ions -- following recent experimental 
and theoretical findings that transition metal dopants provide an enhanced 
carrier mobility which leads to a pronounced increase in conductivity 
with no changes in spectral transmittance \cite{my2prl,Meng,Yoshida-effmass}.

\subsubsection{V.1 Site location of Sn, Ga, Ti or Zr in InGaZnO$_4$}

First, we determine which cation in the host structure is most likely to be substituted 
by the particular dopant. For this, 
we compare the energies of the relaxed structures where In or Zn is substituted 
by Sn or Ga; and In, Ga or Zn is substituted by Ti or Zr. 
Other configurations, e.g., Sn substituted on Ga sites, were not considered 
due to the large ionic radius mismatch.
From the calculations of the defect formation energies,
we find that Sn, Ti and Zr have a strong preference to substitute In atoms
whereas the substitution of Ga or Zn atoms corresponds to a higher energy 
by at least 1.4 eV, 1.3 eV and 3.6 eV, respectively. 
(The formation energies are E$_{form}$(Sn$_{Zn}$)=2.03 eV, 
E$_{form}$(Ti$_{Ga}$)=--1.44 eV, E$_{form}$(Ti$_{Zn}$)=--1.52 eV, 
E$_{form}$(Zr$_{Ga}$)=--0.29 eV, E$_{form}$(Zr$_{Zn}$)=--0.33 eV. These should
to be compared to the formation energies given in Table \ref{subst}).
The obtained preferential substitution can be understood based 
on the local symmetry arguments: ions with larger positive charge 
(M$^{4+}$ in our case) prefer to have more oxygen neighbors and 
so the octahedral oxygen coordination (In sites) 
is more favorable than the tetrahedral one (Zn or Ga sites).
Here we note that the obtained formation energies for doped InGaZnO$_4$ are
similar to those in doped In$_2$O$_3$. For the latter, we find 
the formation energy of the Sn substitution into In site to be equal 
to 0.17 eV (which is in agreement with Ref. \cite{zunger-prl}), 
whereas for the Ti substitution E$_{form}$=--2.92 eV.

In contrast, Ga shows a preference to substitute Zn atoms with the energy 
for In substitution being only 0.05 eV higher (E$_{form}$(Ga$_{In}$)=--0.27 eV
which should be compared to the value for Ga$_{Zn}$, cf., Table \ref{subst}). 
Because of the relatively small energy difference, one can expect that some of the  
extra Ga ions may substitute In atoms upon growth at elevated temperatures. 
This is a disadvantage 
since Ga has the same valence as In and so it will not contribute extra electrons
upon the substitution.  

In Table \ref{subst}, the optimized distances between the dopant 
and its nearest oxygen neighbors are compared 
for doped InGaZnO$_4$ -- along with the calculated electron velocities 
and the optical properties to be discussed in Section V.3.

\subsubsection{V.2 Electronic properties of doped InGaZnO$_4$}

We find that the electron velocity, Table \ref{subst}, is nearly isotropic
in all doped materials -- despite the non-uniform distribution of the dopants 
which prefer to reside within the InO$_{1.5}$ layers.
This is in accord with the hybrid nature of the conduction band in InGaZnO$_4$ 
discussed above (see Section IV) and in our earlier work \cite{my2epl}.
Indeed, in the Sn, Zr and Ga doped cases, the net contributions to the conduction band wave function from the atoms that belong to either In-O or Ga-Zn-O layers are similar, namely,  
58\% or 42\% in Sn case, 51\% or 49\% in Ga and Zr cases, respectively. 
Furthermore, compared to the oxygen deficient material, see Section IV.3, 
the charge density in the conduction band of the Sn, Zr and Ga doped InGaZnO$_4$ is more uniformly distributed  throughout the cell since the largest deviations in the atomic contributions are only a factor of 1.2-1.6, 1.4-1.5, 1.1-1.4, 1.2-1.7 and 1.1-1.6 for, respectively, In, Ga, Zn, O(1) and O(2) located at different distances from the dopant. 
(Note that similar to the oxygen reduced complex oxides of the InGaO$_3$(ZnO)$_n$ family, 
the isotropic behavior may not be maintained in substitutionally doped materials
with the increased number of ZnO layers ($n$) and anisotropic transport properties may appear after annealing.)

The more uniform charge density distribution in the conduction band of Sn, Ga and Zr doped 
InGaZnO$_4$ leads to almost twice larger electron velocities ($\sim$8$\times$10$^5$ $m/s$) than in 
the oxygen deficient material (4$\times$10$^5$ $m/s$) with the concentration of the partially 
compensated oxygen defect of 1.75$\times$10$^{21}$ $cm^{-3}$. (In this case, we compare the same defect concentration, with Sn or V$_O^{+}$ giving one extra electron per impurity or per defect.)
Similar results are obtained for Sn-doped In$_2$O$_3$ where 
the electron velocity is 1.4 times larger (9.17$\times$10$^5$ $m/s$ for 
the dopant concentration of 1.93$\times$10$^{21}$ $cm^{-3}$) as compared 
to the oxygen deficient indium oxide (6.42$\times$10$^5$ $m/s$ for
1.96$\times$10$^{21}$ $cm^{-3}$).
One should take into account the {\it carrier} concentration 
in each case which is determined by the corresponding density of states (DOS) 
in the vicinity of the Fermi level.
The DOS(E$_F$) increases when the energy dispersion of the conduction band 
decreases which represents a more localized state.
The conductivity $\sigma$ depends on both, the electron group velocity
and the density of states near Fermi: 
\begin{equation}
\sigma = \frac{2e^2}{\Omega} \sum_{k\lambda} |v_{k\lambda}|^2 \tau_{k\lambda} \delta (E_{k\lambda}-E_F).
\label{sigma}
\end{equation}
Here $e$ is the electron charge, $\Omega$ -- the volume of the Brillouin zone,
$k$ -- the wave vector, $\lambda$ -- the band index, 
$v$ -- the electron group velocity and $E_F$ is the Fermi energy.
Based on our calculations, we can estimate the band-structure conductivity 
factor -- the square of the electron velocity by the density of states, 
Eq. \ref{sigma}. The estimates suggest that the difference 
between oxygen deficient and Sn-doped oxides, InGaZnO$_4$ and In$_2$O$_3$, 
are almost negligible. Therefore,  
%
the relaxation time for the Fermi surface electrons, $\tau$, 
will play crucial role in determining the resulting conductivity.
While we cannot estimate $\tau$ directly from the band structure calculations, 
below we speculate about the scattering mechanism associated with the Coulomb interaction 
between the ionized donor impurities and free electrons. (We do not consider the effects 
of grain boundaries, structural disorder and neutral impurities -- which may have a significant effect on the transport properties.)

According to \cite{Dingle,Brooks,Bellingham}, the relaxation time 
depends on the concentration of impurity centers, $N_i$, 
which have charge $Ze$ as well as the free electron density $n$.
Because the charge state of the Sn$^{4+}$ substituting In$^{3+}$ ions,
i.e., of the Sn defect, is the same as for the partially compensated 
oxygen vacancy, V$_O^{+}$, there is no apparent difference 
in the relaxation times for the two carrier generation mechanisms 
($Z$=1 and $N_i$=$n$ for both). 
However, we find that the structural relaxation is significantly stronger around the oxygen defect 
than near the substitutional dopant. For In$_2$O$_3$, 
the displacement of the nearest neighbor atoms from their original positions 
is 0.084 \AA \, and uniform around Sn ions, 
while it is much stronger and anisotropic around the partially compensated 
oxygen vacancy with the displacements of 0.081 \AA, 0.112 \AA, 0.127 \AA \, and 0.160 \AA \,
for the four nearest neighbor In atoms. 
For InGaZnO$_4$, the nearest neighbor displacements are 0.069-0.084 \AA \,
in Sn-doped material and 0.097-0.331 \AA \, in the oxygen deficient oxide. 
Qualitatively, the larger atomic displacements around the defect are expected to give rise 
to significantly stronger scattering in the oxygen reduced oxides as compared 
to the substitutionally doped materials.
Most importantly, a strong Coulomb attraction between the V$_O^{+}$ defects 
and the free carriers and, hence, a short electron relaxation time, should be expected
in all oxides considered in this work, since the partially compensated oxygen vacancy corresponds 
to a metastable state. 

Thus, proper doping with aliovalent substitutional impurities is expected 
to lead to a notably higher electron mobility and conductivity than those 
observed in the oxygen reduced oxides.

\subsubsection{V.3 Role of the dopant's electronic configuration}

To elucidate the differences in the calculated electron velocities in 
Ga, Sn, Ti and Zr-doped InGaZnO$_4$, 
cf., Table \ref{subst}, we compared the relative atomic contributions to 
the conduction band wave functions in these materials.
First we find that in the Ga, Sn and Zr cases, the anions give larger net contributions
(64-66 \%) compared to the cations (34-36 \%), while in the case of the Ti doping the cations
contribute twice as much (62 \%) to the total charge in the cell. Moreover, in Ti-doped
InGaZnO$_4$ about 64 \% of 
the net {\it cations'} contribution is due to the Ti $d$-states. In contrast, the states of 
Sn, Ga (primarily $s$-states) or Zr (primarily $d$-states) give only 18\%, 5\% or 7\% of 
the total contributions from all cations in the cell.
Therefore, the extra charge associated with the aliovalent substitution 
is uniformly distributed throughout the cell in Sn, Ga and Zr-doped InGaZnO$_4$, while
it is mostly localized on the $d$-states of Ti \cite{zr} resulting in 
the significantly lower electron velocity in the Ti-doped InGaZnO$_4$.
However, taking into account the increased density of states in the latter case,
the band-structure conductivity factor, $N(\epsilon)v^2(\epsilon)$, cf., Eq. \ref{sigma}, 
is approximately similar for all dopants investigated, 
with Ti showing the largest contribution to the band-structure conductivity, 
Table \ref{subst}.

Further, we investigate how the dopant's electronic configuration
and ionic radius affect the optical properties of InGaZnO$_4$.
In Table \ref{subst}, the fundamental band gap and the Burstein-Moss (BM) shift  \cite{BM}
are compared for doped materials.
The fundamental band gap is measured between the top of the valence
band and the bottom of the conduction band (i.e., at $\Gamma$ point), while the optical band gap
in doped materials corresponds to the energy transitions from the top of the valence band to the Fermi level and, hence, is equal to the fundamental band gap plus the Fermi energy displacement (the BM shift). Note that, in contrast to oxygen deficient InGaZnO$_4$, the substitutionally doped material does not exhibit a second gap, as expected from hybridization between the dopant's states with the $p$ states of neighboring oxygen atoms.

We find that the BM shift is $\sim$0.4 eV smaller for the Ti-doped InGaZnO$_4$ 
as compared to Ga, Sn or Zr, Table \ref{subst}, although Ti$^{4+}$ provides 
the same number of extra electrons per substitution of In$^{3+}$ as Sn$^{4+}$ or Zr$^{4+}$
(or as the substitution of Zn$^{2+}$ with Ga$^{3+}$).
This is explained by the significant charge localization on the Ti $d$-states 
located in the vicinity of the Fermi level \cite{zr}.
The smaller BM shift should not affect
the optical transparency within the visible range: because the band gap in InGaZnO$_4$ 
is large enough, 3.5 eV \cite{Orita-exp}, the energy of the transitions from the valence band 
is above the visible range (i.e., above 3.1 eV) even in the undoped oxide.

Moreover, we argue that a smaller BM shift associated with the electron localization on the dopant $d$-states allows one to balance the optical and transport properties more efficiently \cite{my2prl}.
In a conventional TCO, e.g., Sn-doped In$_2$O$_3$, an increased carrier concentration
desired for an improved conductivity comes at a cost of an increased optical absorption 
because a pronounced BM shift reduces the energy of the transitions from the Fermi level, 
i.e., from the partially occupied conduction band. In addition, 
plasma oscillations
may affect the optical properties by reflecting the electromagnetic waves
of frequency below that of the plasmon.
In contrast to the conventional (non-d element) doping, 
transition metal dopants may allow one to introduce large carrier concentrations without compromising on the optical transparency. 
Indeed, smaller BM shift in Ti-doped InGaZnO$_4$ implies that the impurity 
concentration can be increased further (up to as high as 
1.22$\cdot$10$^{22}$cm$^{-3}$, based on our electronic band structure
estimates) before reaching the same energy of the optical transitions 
from the Fermi level as with Ga, Sn or Zr doping at the dopant concentration 
of 1.75$\cdot$10$^{21}$cm$^{-3}$. We stress here that the above estimates are
only to emphasize the trend between the $s$- and $d$-dopants.
In a viable TCO, the impurity concentration is limited by ionized impurity 
scattering on the electron donors as well as by the plasma frequency. 

Very recently, the experimental results for Mn doped InGaZnO$_4$ were 
reported \cite{Liu-IGZO-Mn}. It was found that the conductivity decreases 
while the carrier mobility increases with Mn content (up to 2.5 at.\% of Mn). 
Such an intriguing behavior may be associated with the onset of the long-range 
magnetic interactions between the Mn ions that leads to significantly longer 
relaxation time for the carriers of a particular spin \cite{my2prl}. 
At the same time, the density of states near the Fermi level, and thus, the
{\it carrier} concentration, decreases once the stable magnetic configuration 
is achieved.
Accurate band structure calculations which go beyond LDA to correct the band 
gap value and the position of the localized Mn $d$-states with respect 
to band edges should be performed to verify this supposition.

\begin{table}
\caption{Optimized distances between the interstitial oxygen and 
its nearest cation neighbors, in \AA, 
and the respective total energy $\Delta$E, in meV, for InGaZnO$_4$ systems 
with O$_i$ located in various positions. The division into the groups is made based
on the optimized cartesian $z$ coordinate, see text and Figure \ref{str}.
}
\label{oi-complex}
\begin{center}
\begin{tabular}{lcccr} \hline
 &  \multicolumn{3}{c}{Nearest neighbor cations} & $\Delta$E \\ \hline 
               
InO$_{1.5}$ layer   &&&& \\
\, 1 $z$=3.1 & \, Ga 1.87 \, & \, In 2.14 \, & \, In 2.99 \, &   0  \\
\, 2 $z$=3.1 & Ga 1.88 & In 2.15 & In 2.92 &  +22 \\
\, 3 $z$=3.0 & Zn 1.92 & In 2.11 & In 2.91 &  +11 \\

GaZnO$_{2.5}$ layers &&&& \\ 
\, 4 $z$=9.4 & Ga 1.85 & Ga 2.04 & Zn 2.46 & +112 \\
\, 5 $z$=9.3 & Zn 1.91 & Zn 2.03 & Ga 2.31 & +253 \\
\, 6 $z$=7.3 & Zn 1.90 & Zn 1.96 & Ga 2.49 &  +72 \\

In between/shared   &&&&  \\
\, 7 $z$=5.1 & Ga 1.87 & Zn 1.89 & Zn 1.90 & +1,490 \\ 
\, 8 $z$=4.7 & Ga 1.84 & In 2.33 & Zn 2.43 & +127 \\
\, 9 $z$=5.5 & Ga 1.89 & Ga 1.89 & In 2.91 & +447 \\ \hline
\end{tabular}
\end{center}
\end{table}

\subsubsection{V.4 Charge compensation with interstitial oxygen}

It is known that high oxygen pressure can significantly reduce 
the conductivity of doped oxides due to the formation of neutral 
complexes between the ionized dopants and the interstitial 
oxygen atoms \cite{Wit,Frank,Gonzalez}.
Indeed, it was shown \cite{my2prl} that interstitial oxygen in Mo-doped
In$_2$O$_3$ may significantly affect the electronic, magnetic and optical
properties of the material. However, the formation energy of the O$_i$ defect 
in undoped In$_2$O$_3$ is very high (we obtained E$_{form}$=8.88 eV for 
the oxygen-rich conditions that is in agreement with Ref. \cite{zunger-prl}).
For a dopant--interstitial-oxygen defect complex 
(e.g., Mo$_{In}^{\bullet \bullet \bullet}$O$^{''}_i$)$^{\bullet}$ -- given
in Kr\"{o}ger-Vink notations)
the energy of formation is lower by about 3 eV as compared 
to the interstitial oxygen alone \cite{my2prl}. 

Given its compositional and structural complexity, InGaZnO$_4$ is a good 
test structure for investigation of the charge compensation mechanism 
via oxygen interstitial. 
Significantly, we find that the formation energy of the neutral O$^{''}_i$
defect in the multicomponent oxide
is 4.77 eV which is almost twice lower than that in In$_2$O$_3$.
This suggests that the formation of dopant-O$_i$ defect complexes in
InGaZnO$_4$ is possible. While studies of particular dopants and 
defect complexes in InGaZnO$_4$ is beyond the scope of this work, 
here we would like to understand the preferred distribution 
of the interstitial oxygen atoms in the complex oxide with chemically and 
structurally distinct layers. For this, we performed calculations for 
the undoped InGaZnO$_4$ with extra oxygen atom located in various structurally 
non-equivalent sites in the lattice.
We find nine such positions for O$_i$ which we divide into three groups 
according to their optimized Cartesian $z$ coordinate: 
(i) the first three of them (cases 1, 2 and 3 in Table \ref{oi-complex})
have $z$=3.0--3.1 which is slightly above the framework oxygen atoms in the InO$_{1.5}$ layer with
$z$=1.9--2.2, cf., Fig. \ref{str}; 
(ii) another three with $z$=9.3 (cases 4 and 5) and $z$=7.3 (case 6) are within 
the GaZnO$_{2.5}$ double layer where the framework oxygen atoms have $z$=6.2--6.5 and $z$=9.9--10.2;
(iii) the remaining three with $z$=4.7--5.5 (cases 7, 8 and 9) reside in the ``shared'' space between 
the structurally and chemically different layers, and are about 2.5 times closer 
to the GaZnO$_{2.5}$ layer.

Table \ref{oi-complex} lists the respective total energies and the optimized distances 
between the interstitial oxygen and its nearest cations for the structures with different 
location of the defect. As one can see, the interstitial oxygen prefers to 
reside within the InO$_{1.5}$ layer having the distance of $\sim$2.1 \AA \, 
from one of the In atoms 
(cases 1, 2 and 3). The total energy deviates insignificantly when 
the next nearest neighbor is either Ga or Zn.
When the distance between the O$_i$ and In atom increases (to 2.3, 2.9 and 3.2 \AA) 
so does the total energy (by 127, 447 and 1,490 meV, respectively).
The energy decreases, however, when the interstitial oxygen atoms 
reside within the GaZnO$_{2.5}$ layers. Nonetheless, the energy difference 
between the structures with O$_i$ located within the InO$_{1.5}$ or GaZnO$_{2.5}$
layers is at least 50 meV. Because the substitutional dopants also have 
a preference to reside within the InO$_{1.5}$ layers, as shown above, 
the formation of the charge compensated complexes is likely.
Therefore, the ambient oxygen pressure must be carefully controlled to avoid 
the charge compensation and to ensure that the largest carrier concentration 
per substitutional dopant is attained. Using dopants which donate more 
than one electron per substitution, such as Mo$^{6+}$ \cite{my2prl}, 
represents an alternative route to mitigate the charge compensation and, thus, 
to achieve good electrical conductivity.

\subsection{VI. Conclusions}

The results of our comparative electronic band structure investigations of 
In$_2$O$_3$, ZnO, Ga$_2$O$_3$ and InGaZnO$_4$ allow us to draw several 
important conclusions:

1) The electronic and optical properties of the oxides originate from 
the strong interaction between the metal $s$-orbitals and the $p$-orbitals of the neighboring oxygen atoms.
Due to the M$s$-O$p$ overlap, the undoped stoichiometric oxides exhibit 
large band gaps (3.4-4.9 eV) and small electron effective masses 
(0.28-0.35 m$_e$). The latter, however, cannot explain the differences 
in the observed electrical conductivities in degenerately doped oxides.

2) Due to the three-dimensional M$s$-O$p$ network, all undoped stoichiometric oxides, including those with well-defined {\it crystal lattice anisotropy}, are capable to give rise to (nearly) isotropic conductivity. In layered InGaZnO$_4$, the electron effective masses are the same within and across the structurally and chemically distinct layers, as confirmed by our sX-LDA calculations.

However, long-range structural anisotropy 
favors non-uniform (preferential) distribution of carrier donors. 
In $\beta$-Ga$_2$O$_3$, oxygen vacancies are most likely to appear at the O(1) sites which form zig-zag chains along the $y$ axis leading to a twice larger electron velocity in this direction.  
In InGaZnO$_4$, an oxygen vacancy, interstitial oxygen atom and substitutional Sn, Zr and Ti prefer to reside within the InO$_{1.5}$ layer that may explain the observed temperature-dependent anisotropy of the electrical conductivity in InGaO$_3$(ZnO)$_n$ and In$_2$O$_3$(ZnO)$_n$ with an increased number of ZnO layers, $n$.

3) {\it Octahedral oxygen coordination} provides the largest $sp$ overlap, 
giving rise to a 1.6 and 2.5 times larger electron velocity in hypothetical 
oxygen deficient rock salt ZnO and $\alpha$-Ga$_2$O$_3$ as compared to 
wurtzite ZnO and $\beta$-Ga$_2$O$_3$, respectively.
Therefore, a way to improve conductivity and to utilize the potential of 
an oxide is to ensure that the cations and anions possess a high {\it local} 
symmetry which can be attained  
via stabilization of a higher-symmetry phase or by choosing a multicomponent 
material with, preferably, octahedral oxygen coordination for all constituents.

4) Most significantly, we find that the electronic configuration of cation(s), 
in particular, the energy proximity of its {\it empty $p$-states with respect 
to the conduction band minimum} (CBM), plays the key role in determining the 
transport properties of oxygen deficient materials. In $\beta$-Ga$_2$O$_3$, 
the Ga $p$ states 
are energetically compatible with the $s$ states of the host cations and, 
thus, available for the vacancy-induced electrons. As a result, in marked 
contrast to In$_2$O$_3$ and ZnO, the carriers become trapped on the $p$-states
of the Ga atoms nearest to the defect -- owing to the weak hybridization 
of these highly anisotropic Ga $p$-orbitals with the $p$-orbitals of their 
neighboring oxygen atoms.

This finding explains why $\beta$-Ga$_2$O$_3$ is not a viable TCO and 
also suggests that a large content of Ga$_2$O$_3$ in a multicomponent TCO 
may lead to a suppressed conductivity -- unless an alternative 
carrier generation mechanism is found.

5) Due to the hybrid nature of the conduction band, multicomponent oxides offer a possibility to overcome the detrimental electron localization effects
by tuning the resulting electronic -- as well as optical -- 
properties via chemical composition. 

6) In all oxides investigated in this work, neutral oxygen vacancy, 
V$_O^0$, results in a fully occupied, thus, non-conducting 
state located well below the oxide conduction band minimum. 
%
Because V$_O^0$ also corresponds to the ground state of 
the defect, a conducting behavior may appear only if the vacancy is excited 
to V$_O^{+}$.
In In$_2$O$_3$, ZnO and InGaZnO$_4$, our LDA estimates suggest that 
the energy of excitation is within the visible range, i.e., the vacancy can 
be photo-excited, leading to a metastable conductive state. 
In $\beta$-Ga$_2$O$_3$, the excitation is highly unlikely because 
V$_O^0$ and V$_O^{+}$ prefer to be located at 
the different oxygen sites - of type O(3) and O(1), respectively. 

7) Aliovalent substitutional doping with next-column elements gives rise 
to twice larger electron velocity as compared to oxygen deficient materials. 
However, the carrier concentration is expected to be lower in the former 
due to a smaller (by order(s) of magnitude) density of states in the vicinity 
of the Fermi level. This is associated 
with a nearly rigid-band shift of the Fermi level upon substitutional doping, 
whereas notable electron localization near the oxygen vacancy leads to
a higher DOS near the Fermi level in oxygen deficient materials.

Our results suggest that the differences in the conductivities in oxygen 
deficient and doped oxides stem from different carrier relaxation time. 
A shorter $\tau$ in oxygen deficient oxides is expected from a significant 
atomic relaxation around the oxygen defect and a stronger Coulomb attraction 
between the oxygen vacancy V$_O^{+}$ and the free carriers.


%

8) Doping with specific transition metal elements represents 
an efficient way to achieve high electrical 
conductivity without substantial changes in the spectral transmittance. 
In addition, it provides a possibility to tune the electronic and
optical properties of oxide via the formation of charged defect complexes 
with interstitial oxygen atom. 

Thus, in-depth understanding of the microscopic properties of 
the post-transition metal oxides provides significant insights into 
the underlying phenomena in the conventional TCOs. Moreover, our systematic studies reveal several ways to knowledgeably manipulate the resulting properties - by tuning the chemical composition of complex hosts and carefully choosing proper carrier generation mechanisms - to give rise to broader application range of these unique materials.

\subsection{Acknowledgment}

The work is supported by 
the National Science Foundation (NSF) (grant DMR-0705626) and
the Petroleum Research Fund of the American Chemical Society (grant 47491-G10).
Computational resources are provided by the NSF supported TeraGrid
and the National Energy Research Scientific Computing 
Center (NERSC) which is supported by the Office of Science of 
the U.S. Department of Energy under contract DE-AC02-05CH11231.




\begin{thebibliography}{99}
\bibitem{mysXLDA}
The results of the screened-exchange LDA will be published elewhere.
\bibitem{zr}
Note that in contrast to Ti-doped InGaZnO$_4$ where the Ti $d$-states are located 
0.15 eV above the Fermi level, the Zr $d$-states are significantly higher in energy, 
namely, $\sim$1.0 eV above the Fermi level, and thus are not available for 
the extra electrons.
\end{thebibliography}
\end{document}